\documentclass[preprint]{aastex}
\usepackage{graphics}
\usepackage{epsfig}
\usepackage{ifthen}
\usepackage{natbib}

\def\kms{\rm\,km\,s^{-1}}

\def\cm{{\rm\,cm}}

\def\erg{{\rm\,erg}}
\def\kev{{\rm\,keV}}

\def\ev{{\rm\,eV}}
\def\kms{{\rm\,km\,s^{-1}}}
\def\K{{\rm\,K}}
\def\lya{{Ly$\alpha$}}
\def\hmpc{{h^{-1}{\rm Mpc}}}
\def\hkpc{{h^{-1}{\rm kpc}}}
\def\spose#1{\hbox to 0pt{#1\hss}}

\def\ovi{{\ion{O}{6}}}
\def\ovii{{\ion{O}{7}}}
\def\oviii{{\ion{O}{8}}}
\def\oix{{\ion{O}{9}}}
\def\neix{{\ion{Ne}{9}}}
\newcommand{\Msun}{\ensuremath{M_\odot}}
\newcommand{\Mpc}{\ensuremath{\mathrm{Mpc}}}
\newcommand{\keV}{\ensuremath{\mathrm{keV}}}
\newcommand{\Sec}{\ensuremath{\mathrm{s}}}

\newcommand{\sr}{\ensuremath{\mathrm{sr}}}
\newcommand{\Hz}{\ensuremath{\mathrm{Hz}}}
\newcommand{\dd}{\ensuremath{\mathrm{d}}}
\newcommand{\mang}{\ensuremath{\hbox{m\AA}}}

\begin{document}

\title{X-ray Absorption by the Low-redshift Intergalactic Medium:\\
A Numerical Study of the $\Lambda$CDM model}

\author{Xuelei Chen\altaffilmark{1}}

\affil{Department of Physics, The Ohio State University, Columbus,
OH 43210, USA}
%\email{xuelei@itp.ucsb.edu}

\author{David H. Weinberg}
\affil{Department of Astronomy, The Ohio State University, Columbus,
OH 43210, USA}
%\email{dhw@astronomy.ohio-state.edu}

\author{Neal Katz}
\affil{Department of Astronomy, University of Massachusetts,
Amherst, MA 01003, USA}
%\email{nsk@kaka.astro.umass.edu}

\author{Romeel Dav\'{e}\altaffilmark{2}}
\affil{Steward Observatory, University of Arizona, Tucson, AZ 85721, USA}
%\email{rad@as.arizona.edu}

\altaffiltext{1}{Present address: Institute for Theoretical Physics,
U.C. Santa Barbara, Santa Barbara, CA 93106, USA}

\altaffiltext{2}{Hubble Fellow}

\begin{abstract}
Using a hydrodynamic simulation of a cold dark matter universe with a
cosmological constant ($\Lambda$CDM), we investigate the ``X-ray forest''
absorption imprinted on the spectra of background quasars by the intervening
intergalactic medium (IGM), at redshift $z\approx 0$.
In agreement with previous studies, we find that
\ovii\ and \oviii\ produce the strongest absorption features.  The strong
oxygen absorbers that might be detectable with {\it Chandra} or {\it 
XMM-Newton}
arise in gas with $T \sim 10^{5.5}\K - 10^{6.5}\K$ and overdensities
$\delta \ga 100$ that are characteristic of galaxy groups.  Future X-ray
missions could detect weaker oxygen absorption produced by gas with a wider
range of temperatures and the lower densities of unvirialized structures; they
could also detect X-ray forest absorption by carbon, nitrogen, neon, iron, and
possibly silicon.  If the IGM metallicity is $Z=0.1Z_\odot$, as we assume in
most of our calculations, then the predicted number of systems strong enough
for a $\sim 5\sigma$ detection with {\it Chandra} or {\it XMM-Newton} is
extremely low. However, scatter in metallicity increases the number of 
strong absorbers even if the mean metallicity remains the same, 
making the predictions somewhat more optimistic.
Our simulation
reproduces the high observed incidence of \ovi\ (1032\AA, 1038\AA) absorbers,
and the most promising strategy for finding the X-ray forest is to search at
the redshifts of known \ovi\ systems, thus reducing the signal-to-noise
threshold required for a significant detection.  However, while many \ovi\
absorbers have associated \ovii\ or \oviii\ absorption, the \ovi\ systems
trace only the low temperature phases of the X-ray forest, and a full 
accounting
of the strong \ovii\ and \oviii\ systems will require a mission with the
anticipated capabilities of {\it Constellation-X}.  The large effective area
of the {\it XEUS} satellite would make it an extremely powerful instrument for
studying the IGM, measuring X-ray forest absorption by a variety of elements
and revealing the shock-heated filaments that may be an important reservoir
of cosmic baryons.
\end{abstract}


\section{Introduction}

The primordial deuterium abundance, combined with the theory of
big bang nucleosynthesis, implies a cosmic baryon density
$\Omega_b \approx 0.02 h^{-2}$ (\citealt{burles97,burles98};
here $h\equiv H_0/100\;\kms\Mpc^{-1}$), in agreement with
recent estimates from the cosmic microwave background anisotropy
measurements \citep{deBernardis01,spergel03}.
This value is an order of magnitude
higher than the estimated density of stellar mass
(e.g., \citealt{FHP98}).  At redshifts $z\sim 2-4$, it appears
that most baryons reside in the diffuse, photoionized
intergalactic medium (IGM) that produces the \lya\ forest:
that is what hydrodynamic simulations predict
\citep{CMOR94,ZAN95,HKWM96,MCOR96,theuns98}, and it is what the measured
opacity of the \lya\ forest implies given reasonable estimates of the
neutral fraction \citep{rauch95,HKWM96,rauch97,weinberg97,schaye01}.
By low redshift, according to the simulations, some of this gas
makes its way into galaxies and some remains in a diffuse
photoionized medium at $T\sim 10^4\K$, but a significant fraction
is shock-heated to higher temperatures \citep{CO99a,DHKW99}.
\citet{CO99a} coined the term ``warm-hot intergalactic medium''
(WHIM) to refer to intergalactic gas in the temperature range
$10^5\K < T < 10^7\K$, and a variety of numerical simulations
predict that 30-40\% of baryons reside in the WHIM at $z=0$
\citep{dave01}.

While the hot ($T\sim 10^7-10^8\K$), dense
($\delta \equiv \rho/{\bar \rho_b} > 100$) gas in galaxy clusters
can be readily detected in X-rays, most of the shocked IGM lies at
lower temperatures and densities, where its X-ray emission is relatively
feeble.  One of the few prospects for observing this component is
via the X-ray {\it absorption} that its highly ionized metal lines
would produce in the spectra of background quasars.
The absence of continuous ``X-ray Gunn-Peterson'' absorption
provided early limits on the density and temperature of the IGM
\citep{SB80,AEMF94}. Anticipating the high spectroscopic resolution
of Chandra and XMM-Newton, and of future, more powerful X-ray
telescopes like Constellation-X and XEUS, \citet[hereafter HGM]{HGM98}
and \citet{PL98} proposed the concept of an ``X-ray forest,'' a
pattern of resonant metal absorption lines analogous to the \lya\
forest.  Using a hydrodynamic cosmological simulation, HGM
predicted distribution functions of \ion{O}{7} and \ion{O}{8}
absorption.  \citet{PL98} and \citet{FC00} used analytic methods
based on the Press-Schechter (\citeyear{press74}) mass function
to make similar predictions, considering additional ionic species and
a range of cosmological models.
X-ray afterglows of $\gamma-$ray bursts could also be used as
background sources, in addition to bright quasars
\citep{FNSSV00}.

This paper examines the X-ray forest phenomenon in a large,
smoothed particle hydrodynamics (SPH) simulation of the $\Lambda$CDM
model (inflationary cold dark matter with a cosmological constant).
We extend the work of HGM by considering a wider range of ionic
species, looking more closely at the physical state of the gas
responsible for X-ray absorption, and examining the prospects for
detection of the X-ray forest with current and next-generation
X-ray telescopes.  Our study is similar in approach to the recent
work of \citet[hereafter FBC]{fang02}, but there are significant differences 
in the atomic processes incorporated in the calculations
(see \S\ref{sec:ionfrac}) and differences in focus of the analysis.

The recent detection of a high density of \ovi (1032\AA, 1038\AA)
absorbers \citep{TS00,TSJ00,S02} offers a tantalizing
hint of a baryon component that might correspond to the warm-hot IGM.
Although we have not designed our study primarily for \ovi\ predictions,
we do calculate the expected incidence of \ovi\ absorption and 
investigate the connections between \ovi\ absorbers and
X-ray forest systems (\ovii\ and \oviii).  \citet{Cen01}
and \citet{fang01} provide
detailed \ovi\ predictions using hydrodynamic simulations.

The next section describes our SPH simulation and our methods
for calculating X-ray forest absorption.
Section 3 presents our predictions for the equivalent width
distributions of a variety of metal lines and discusses the
feasibility of detection.  Section 4 focuses on the physical
properties of the strong oxygen absorbers, which appear to be the
most promising targets for the near future.
We summarize our results in \S 5.

\section{Modeling Methods}
\label{sec:method}

\subsection{Cosmological simulation}
The cosmological simulation analyzed here is identical to the simulation D1
described in \citet{dave01}.
Some illustrations of structure in this simulation appear in
Figures~\ref{zoombox} and~\ref{boxmap}, which are discussed in
\S\ref{sec:physics} below.
We adopt a flat $\Lambda$CDM model with $\Omega_m=0.4$,
$\Omega_{\Lambda}=0.6$, $h=0.65$, $\Omega_b=0.0473$, and the primordial
power spectrum predicted by the inflationary CDM model with an
inflationary spectral index $n=0.95$ and normalization $\sigma_8 = 0.8$.
This cosmological model is consistent with observational constraints
from the cosmic microwave background, large scale structure, galaxy
clusters, the \lya\ forest, and Type Ia supernovae,
though the value of $\Omega_m$ is somewhat higher than favored
by the most recent measurements \citep{spergel03}.
The simulation box is a periodic cube of side-length $50h^{-1}$ comoving Mpc.
At the start of the simulation, it contains $144^3$ dark matter particles
and an equal number of gas particles; some of the gas particles are
converted to collisionless star particles during the course of the
calculation.
The masses of individual dark matter and gas particles are
$6.3\times 10^9\Msun$ and $8.5\times 10^8\Msun$, respectively.
The simulation is evolved from $z=49$ to $z=0$ using the parallel
version \citep{DDH97} of TreeSPH \citep{hernquist89,KWH96}, which utilizes a
tree algorithm for gravitational forces and SPH for hydrodynamic
interactions.  The gravitational force resolution is $7\hkpc$
(equivalent Plummer softening), in comoving units.
The spatial resolution of the hydrodynamic calculation varies with
density because of the Lagrangian nature of the SPH algorithm; all
gas dynamical quantities are computed by smoothing with a symmetric
spline kernel that encloses 32 SPH particles, or at most 
$2.7\times 10^{10}\Msun$
of baryons, with a minimum SPH smoothing length of $2.5\hkpc$ (comoving).

The simulation incorporates shock heating and Compton and radiative
cooling, as well as adiabatic heating and cooling.
Radiative cooling rates are computed for a primordial composition gas,
as described by \citet{KWH96}.
Cold gas in regions with density comparable to the Galactic interstellar
medium is converted into stars; since this has little impact on the IGM
at the densities and temperatures relevant to the X-ray forest, we
refer the reader to \citet{KWH96} for a discussion of the star formation
algorithm.
At the mass resolution of this simulation, inclusion of heating by
a photoionizing background artificially suppresses the formation
of galaxies \citep{WHK97}, so the simulation was run without a
photoionizing background.  As a consequence, expansion of the universe
cools unshocked gas to very low temperatures (less than $100\K$).
In higher resolution simulations that incorporate a photoionizing
background \citep{DHKW99}, unshocked gas at redshift $z=0$ follows a tight
temperature-density relation, $T \approx 10^4(\rho/{\bar\rho}_b)^{0.6}\K$,
which arises from the competition between photoionization heating
and adiabatic cooling \citep{hui97}.  (The mean temperature for this
cosmological model is higher than that in the
\citeauthor{DHKW99} [\citeyear{DHKW99}] simulations
because of the higher baryon density.)
Before analyzing the simulation, therefore, we 
raise the temperatures of
all gas particles
with baryon overdensity $\delta_b < 100$ and 
$T<10^4 \delta_b^{0.6}\K$ so that they 
lie on this temperature-density relation.
We do not adjust the temperatures of shock-heated gas particles that
already lie above this relation.
This after-the-fact adjustment is a good approximation to
the gas temperatures we would obtain in a simulation with an
ionizing background.  Most of the significant X-ray forest absorption
will come from hotter, denser gas that is unaffected by this adjustment.

In this paper, we concentrate entirely on the $z=0$ output of the
simulation.  Owing to the difficulty of detecting X-ray forest
lines at all (as discussed in \S\ref{sec:predictions} below),
observational studies are likely to focus on the quasars with the
highest apparent X-ray fluxes, and hence on low redshift.
We obtain density, temperature, and peculiar velocity profiles
along 1200 randomly chosen lines of sight through the simulation cube,
400 in each projection, using the algorithm described by
\citet{HKWM96}, as implemented in Katz \& Quinn's code
TIPSY.\footnote{
{\tt http://www-hpcc.astro.washington.edu/tools/tipsy/tipsy.html}}
Each line of sight is divided into 5000 bins, so that each bin has a
width of 1 km/s in velocity units, which is much finer than the few
hundred km/s spectral resolution typical of the current and next
generation X-ray telescopes.  The contribution of each gas particle
to each bin of the profile is calculated by a line integral through
its spherically symmetric spline kernel, with contributions going
to zero beyond the smoothing scale that encloses 32 SPH neighbors.
We use the densities and temperatures to compute ionic species
abundances as described below (\S\ref{sec:ionfrac}), then calculate
absorption spectra in redshift space by accounting for peculiar
velocities and thermal broadening.  
We treat the projection axis of the cube as the redshift direction
and make use of the periodic boundary, so a feature that would
be ``shifted out'' of the cube by its peculiar velocity
reappears at the opposite side.
Except for the computation of
ionic abundances, the calculation is entirely analogous to the
calculation of \lya\ forest absorption by \citet{HKWM96}.

\subsection{Selection of transition lines}

Consider an absorption system of a given temperature, density, and
path length.  The strength of the absorption line produced by a given
transition of a given ion is proportional to the path length,
the transition oscillator strength,
the abundance of the element in question, and the fraction of that element
in the given ionization state, which depends in turn on the temperature,
density, and the ionizing background radiation.
The predicted number of lines per unit redshift also depends on the
covering factor of absorbers with the temperature, density, and path length
required to yield detectable absorption.
To decide which lines to consider in our study, we first made a
preliminary screening of all of the ions and transition lines listed in
\citet{transitionlist}. We selected those
with element abundance greater than $10^{-7}$ and
resonance lines with energy above 0.25 keV ($\lambda < 50$\AA).
For each ion we selected the resonance line with the greatest oscillator
strength. We then calculated the total absorption produced along
all 1200 lines of sight through the simulation box, i.e., the summed
equivalent widths of all of the absorption lines produced by these
transitions, and ranked the ions accordingly.
The ions that produced at least one of the ten strongest lines are listed
in Table~\ref{ions}.  In addition to \ion{O}{7} and \ion{O}{8}, we found that
\ion{C}{5}, \ion{C}{6}, \ion{N}{6}, \ion{N}{7}, \ion{Ne}{9},
\ion{Ne}{10}, \ion{Si}{13}, and \ion{Fe}{17} produce relatively
large amounts of absorption.

Detection of X-ray forest lines requires that the observed energy
be larger than $\sim 0.2\;\kev$, both because of instrumental sensitivity
and to avoid absorption by the Galactic interstellar medium.
Lines in Table~\ref{ions} with energy $E$ can therefore be observed
only to a maximum redshift $z=E/(0.2\;\kev)-1$.
\cite{FC00} have used Press-Schechter models to predict absorption
by \oviii\ and by two ionic species with much higher transition energies,
\ion{Si}{14} ($2.01\;\kev$) and \ion{Fe}{25} ($6.70\;\kev$), which
can potentially be observed out to high redshifts (see also \citealt{PL98}).
These last two ionization states become abundant only at the high
temperatures ($T \sim 10^7\K$ and $T\sim 10^8\K$, respectively)
characteristic of rich galaxy clusters, which are poorly represented
in our simulation box owing to its limited size.
We therefore do not attempt to make predictions for these ions.
The analytic method used by \cite{FC00} is well adapted to the study
of these highly ionized species, which are likely to occur only in
the virialized regions that are best described by the Press-Schechter
approximation and which are too rare to be well represented in our
hydrodynamic simulation.  Our numerical approach complements this
analytic method by allowing more accurate investigation of lower
ionization species, which may be found in lower density,
unvirialized regions.

\subsection{Metal abundance}

The \ion{C}{4} and \ion{O}{6} absorption associated with \lya\
forest lines implies that the metallicity of the low density IGM
is $\sim 10^{-3} - 10^{-2}$ of solar at high redshift
(see, e.g., \citealt{SC96,haehnelt96,hellsten97,dave98,ellison00,schaye00}).
At low redshift ($z \simeq 0.5$), \cite{BT98} found [C/H] $\simeq -1.6$ for 
HI
absorbers with $W_r \sim 0.5$\AA\ using Hubble FOS data.
In the denser IGM of X-ray clusters, on the other hand,
the inferred metal abundance is as high as one third of solar
\citep{clusterabundance}.  Unfortunately, these observations give
little guidance to the metallicity of the regions with $\delta \sim 10-1000$
that are responsible for the X-ray forest at $z=0$.
While metal enrichment can be tracked
in a simulation either during dynamical evolution, as in HGM
and \cite{Cen01}, or in post-processing, as in \cite{aguirre01},
the numerical predictions are sensitive to the uncertain
details of metal ejection from the immediate vicinity of galaxies.
Rather than adopt an explicit but model-dependent prescription
for the metallicity
as a function of environment, we will generally assume that the
metallicity of the medium that produces X-ray forest absorption
is $0.1 Z_\odot$, intermediate between the \lya\ forest and cluster IGM
values.  We also assume that the relative abundances of
the different elements are similar to those in the Sun; some authors
(e.g., FBC) have adopted enhanced abundances of $\alpha$-elements
(oxygen in particular), characteristic of Type II SN enrichment.
Since most absorbers are unsaturated, predicted equivalent
widths are simply proportional to the assumed metallicity,
but this proportionality breaks down for the strongest absorbers.
We discuss the impact of different metallicity assumptions,
in particular the effects of including a density-dependent metallicity
and scatter about the mean metallicity-density relation \citep{CO99b},
in \S\ref{sec:ewdist} below.

\subsection{The X-ray background}
\label{sec:xrb}

\begin{figure}
\centerline{
\epsfxsize=3.5truein
\epsfbox{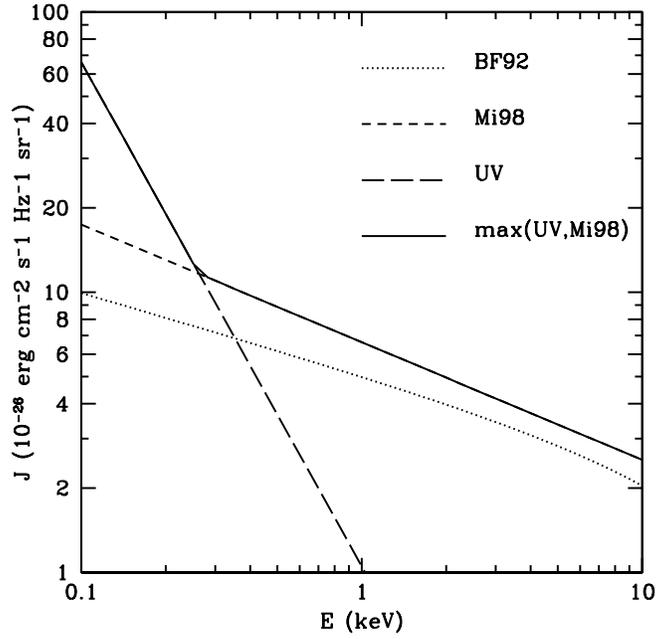}
}
\caption{
The X-ray and UV background spectra used in our calculations.
The dotted line shows the BF92 spectrum, eq.~(\ref{eqn:bf92}).
The short-dashed line, largely obscured by the solid line, shows
the Mi98 spectrum, eq.~(\ref{eqn:mi98}).  The long-dashed line
shows the UV background spectrum of eq.~(\ref{eqn:uvb}),
based on \cite{shull99}.  For most calculations, 
we take the maximum of the UV and Mi98
spectra, as shown by the solid line.  For calculations involving
oxygen ions, we also performed calculations with the BF92 spectrum 
for comparison.
}
\label{XRB}
\end{figure}

As discussed below in \S\ref{sec:ionfrac}, ionic abundances are
affected by both collisional ionization and photoionization.
To compute the latter, we must adopt an intensity and spectrum of
the X-ray background, which contains the ionizing photons that
are relevant to these high ionization species.
For most of our calculations, we adopt the background of
\citet[][hereafter Mi98]{Miyaji98}, which was based on ASCA 
and ROSAT data:
\begin{equation}
J_x(E) = J_1 \left(\frac{E}{\keV}\right)^{-0.42},
\label{eqn:mi98}
\end{equation}
where
$J_1 = 6.626 \times 10^{-26} \erg~ \cm^{-2} \Sec^{-1}\sr^{-1}
\Hz^{-1}$. This has a higher intensity and slightly steeper slope
than the \citet[][hereafter BF92]{BF92} background, which was 
adopted by HGM. The BF92 background spectrum is
\begin{equation}
J_x(E) = J_0 \left(\frac{E}{E_X}\right)^{-0.29} e^{-E/E_X},
\label{eqn:bf92}
\end{equation}
where
$E_X = 40 \keV$ and $J_0 = 1.75 \times 10^{-26} \erg~ \cm^{-2} \Sec^{-1}
\sr^{-1} \Hz^{-1}$.
Figure~\ref{XRB} plots the BF92 and Mi98 background spectra as
dotted and short-dashed lines, respectively.

Extrapolation of these X-ray background fits to longer wavelengths
substantially underestimates the flux expected from quasars
and star-forming galaxies.  While the
UV photoionizing background makes little difference for high ionization
X-ray lines, it does matter for determining the ionic abundances
of HI, \ovi, and, at relatively low densities and temperatures, \ovii.
We therefore include a UV background
\begin{equation}
J_{\rm UV}(E) = J_2 \left(\frac{E}{1\;{\rm Ryd}}\right)^{-1.8},
\label{eqn:uvb}
\end{equation}
with
$J_2 = 2.4 \times 10^{-23} \erg~ \cm^{-2} \Sec^{-1}\sr^{-1} \Hz^{-1}$,
based on \cite{shull99}.
This background is shown by the long-dashed line in Figure~\ref{XRB}.
We generally use the
Mi98 spectrum at energies $E\geq 0.25\;\keV$ and the UV
background~(\ref{eqn:uvb}) at energies $E<0.25\;\keV$, where it
exceeds the Mi98 spectrum, though we have repeated some
of our calculations for different background choices as discussed below.

\subsection{Ionization fractions}
\label{sec:ionfrac}

We calculate the ionization fraction for each ion at a number of
temperature and density grid points with an assumed radiation
background, using
the publicly available code Cloudy \citep{Cloudy}. The ionization
fraction of each ion for any given temperature and density is then
obtained by interpolation.

\begin{figure}
\centerline{
\epsfxsize=5.5truein
\epsfbox{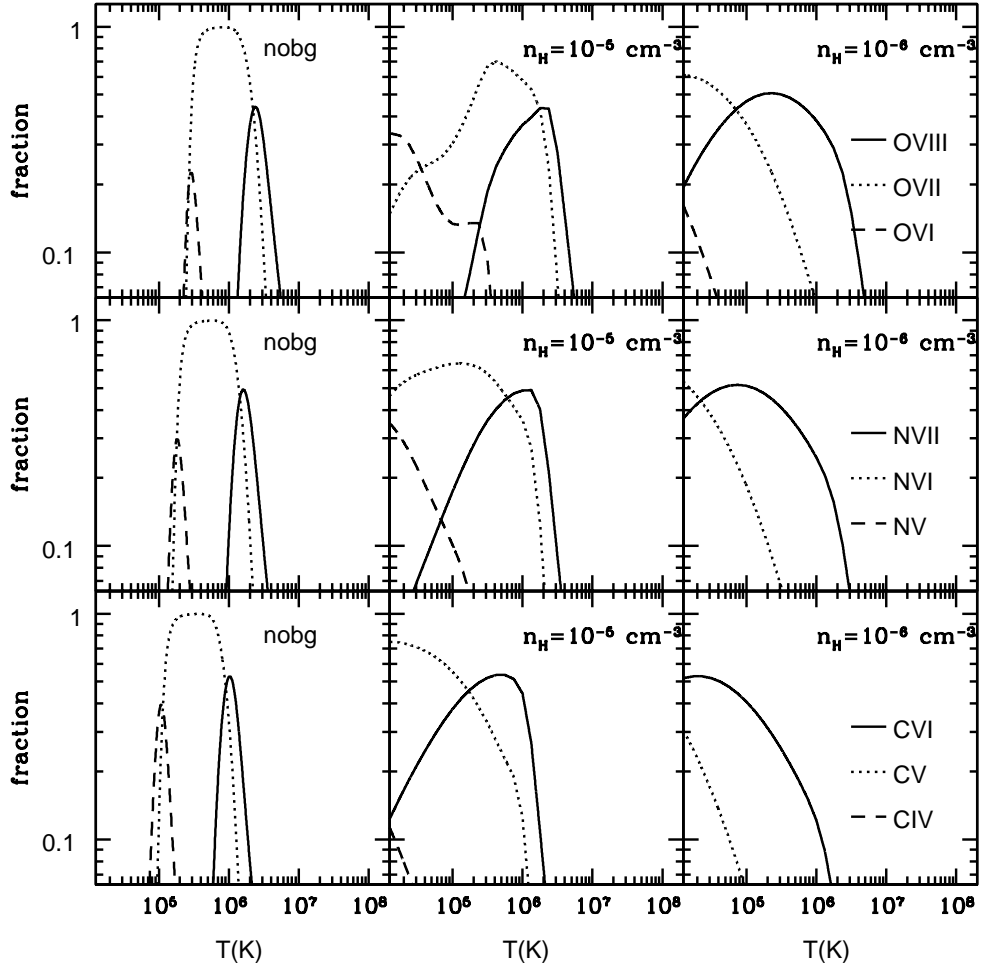}
}
\caption{
Ionization fractions of oxygen (top row), 
nitrogen(middle row), and carbon (bottom row) as a function
of temperature.  Left-hand panels show collisional ionization only.
Central and right-hand panels show ionization fractions including
photoionization by our standard X-ray + UV background
(solid line of Fig.~\ref{XRB}), for densities
$n_H=10^{-5}\cm^{-3}$ ($\delta = 60$) and
$n_h=10^{-6}\cm^{-3}$ ($\delta =6$), respectively.
}
\label{CNOfrac}
\end{figure}

\begin{figure}
\centerline{
\epsfxsize=5.5truein
\epsfbox{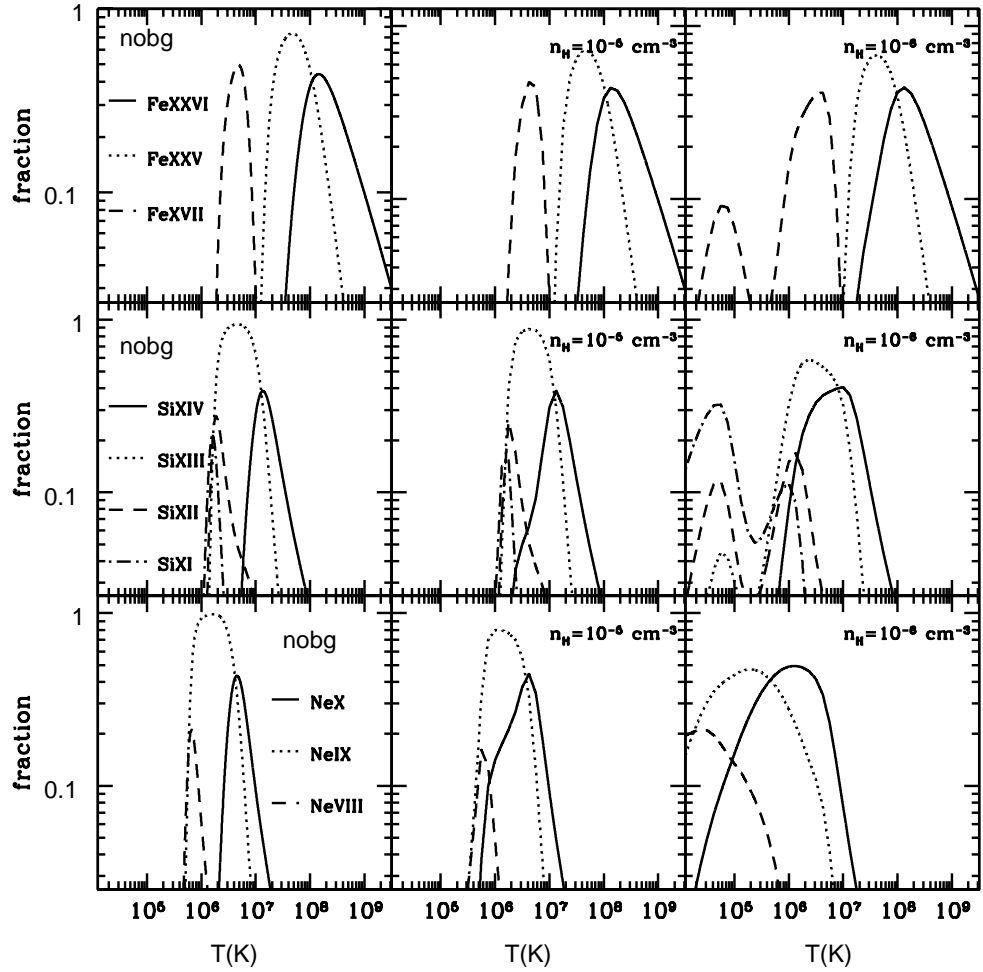}
}
\caption{
Like Fig.~\ref{CNOfrac}, but for iron (top row), silicon (middle row),
and neon (bottom row).  Note the expanded temperature scale 
relative to Fig.~\ref{CNOfrac}.
}
\label{NeSiFefrac}
\end{figure}

Figures~\ref{CNOfrac} and~\ref{NeSiFefrac} illustrate the resulting
ionization fractions that enter our calculations. 
Figure~\ref{CNOfrac} show results for the
hydrogen-like, helium-like, and
lithium-like states of oxygen (top row), nitrogen (middle row),
and carbon (bottom row).
Curves in the central panels are calculated using our 
standard radiation background for a gas density
$n_H=10^{-5}\;\cm^{-3}$, which corresponds to a baryon overdensity
$\delta_b=60$ at $z=0$ for our adopted cosmological parameters.
Right-hand panels show calculations for $n_H=10^{-6}\;\cm^{-3}$,
an overdensity $\delta_b=6$. 
Left-hand panels show results for collisional ionization only,
in which case ionization fractions are independent of density.
(We actually compute these curves using $n_H=10^{-5}\;\cm^{-3}$
and a radiation background reduced in intensity by a factor of $10^4$.)

Since oxygen, nitrogen, and carbon all exhibit similar behavior, we restrict
ourselves to oxygen in the following discussion.
In the absence of photoionization, the
fragile, lithium-like species \ovi\ exists only over a narrow
temperature range, $T \sim 2-5\times 10^5\K$,  and even in this
range its fractional abundance never exceeds 0.25.
The helium-like state, \ovii, dominates a much wider temperature
range, with a fraction exceeding 0.5 for
$T\sim 0.3-2.1\times 10^6\K$.  
At higher temperatures, \ovii\ begins to give way to \oviii,
though the \oviii\ fraction never exceeds $\sim 0.3$ because
temperatures high enough to produce \oviii\ are also high enough
to produce \oix\ (not shown on the plot).
Similarly, at low temperatures \ovii\ and \ovi\ give way to
lower ionization species.

At these densities, photoionization has a major impact on the
predicted abundances.  There is still a general trend of moving
from \ovi\ to \ovii\ to \oviii\ as the temperature increases,
but with photoionization included
each species occupies a much wider temperature range, and there
are more temperatures at which multiple species co-exist.
For $\delta_b=6$, the peak \oviii\ fraction occurs
at $T\sim 2\times 10^5\K$, and the \ovii\ fraction peaks
at $T\sim 10^4\K$.
For $\delta_b=60$, the recombination rates are higher relative to the
photoionization rates, so the curves shift towards those of pure
collisional ionization, but even in this case \ovii\
remains significant down to temperatures of ${\rm several}\times 10^4\K$
and \oviii\ to ${\rm several}\times 10^5\K$.  More generally,
for the densities characteristic of intergalactic filaments or
the outskirts of clusters and groups, photoionization allows a
species to exist at temperatures far below those deduced from
the collisional equilibrium ionization fractions.

The behavior of nitrogen and carbon is entirely analogous to that
of oxygen, except that all the curves are shifted towards lower temperature
owing to the lower ionization energies.  For the more highly
ionized species shown in Figure~\ref{NeSiFefrac}, photoionization
is usually much less significant, since there are fewer high energy
photons.  However, it still makes a difference in some cases
for $\delta_b=6$.  Note the expanded temperature scale of
Figure~\ref{NeSiFefrac} relative to Figure~\ref{CNOfrac}.
The highly ionized iron species, \ion{Fe}{25} and \ion{Fe}{26},
occur only at the high temperatures characteristic of rich
clusters, though \ion{Fe}{17} occurs at the lower temperatures
characteristic of the WHIM.

\begin{figure}
\centerline{
\epsfxsize=5.5truein
\epsfbox{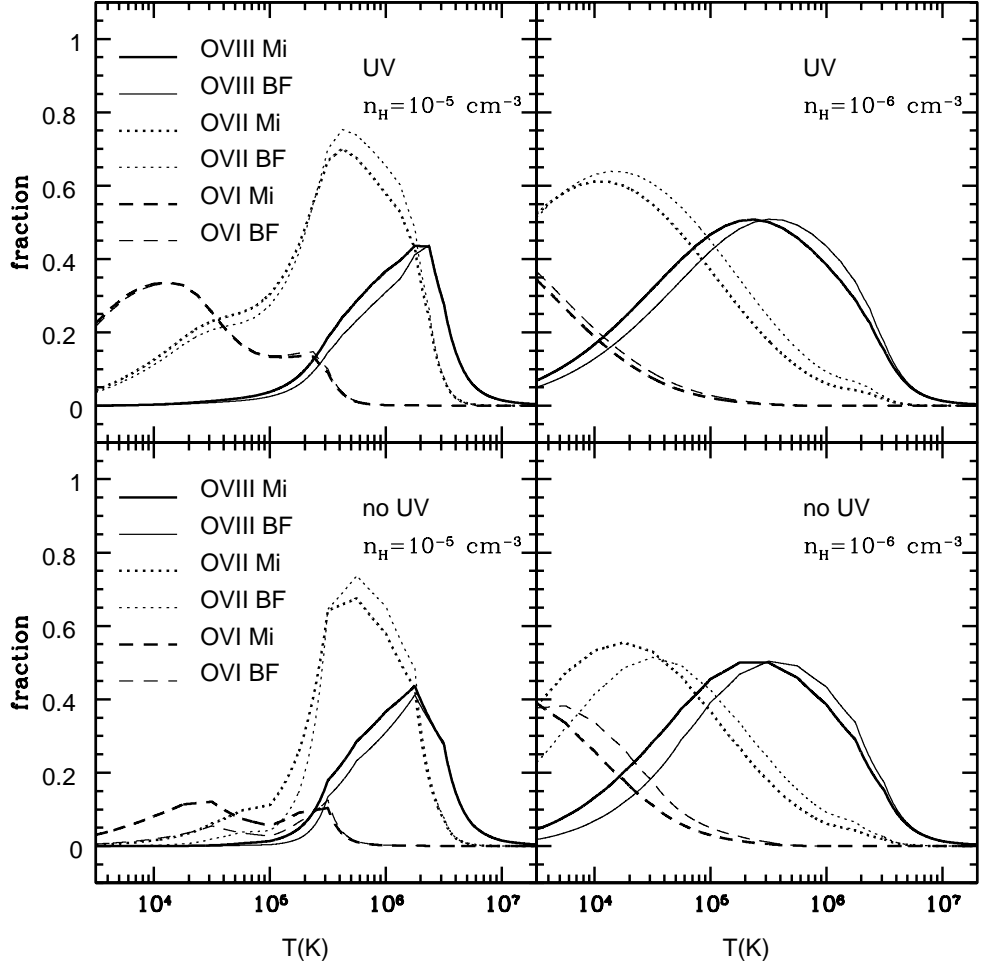}
}
\caption{
Influence of the background radiation on the \ovi, \ovii, and \oviii\ 
fractions.  In the upper panels, heavy curves show our standard 
background, Mi98+UV, as in Fig.~\ref{CNOfrac}, for densities
$n_H=10^{-5}\;\cm^{-3}$ (left) and 
$n_H=10^{-6}\;\cm^{-3}$ (right).  (Relative to Fig.~\ref{CNOfrac},
note the use of a linear vertical scale and the extension to lower
temperatures.)
Light lines show results for the BF92 background instead of Mi98.
Lower panels show the effect of omitting the UV background (but
extrapolating the Mi98 or BF92 spectrum into the UV regime),
which significantly
reduces the \ovi\ and \ovii\ fractions at low temperatures for 
$n_H=10^{-5}\;\cm^{-3}$.
}
\label{BFvsMi}
\end{figure}

Figure~\ref{BFvsMi} shows the dependence of the oxygen predictions
on the adopted radiation background.  In each panel, heavy and
light lines show the ionization fractions for the Mi98 and BF92
X-ray background spectra, respectively.  Upper panels show results
including the UV background spectrum of equation~(\ref{eqn:uvb})
(using the maximum of the X-ray background and UV background at each
wavelength), and lower panels show results using the X-ray background only.
There is little difference between the results that use the Mi98 or BF92 
spectra, indicating that uncertainties in the X-ray background should introduce
little uncertainty in our predictions.  However, the UV background
makes a significant difference for \ovi\ and to some extent \ovii,
particularly at higher densities and lower temperatures.
Uncertainty in the intensity and spectral shape
of the UV background introduces some uncertainty in our \ovi\ predictions,
but it should make little difference to our predictions of X-ray
absorption.

It is worth noting at this point the two main physical differences
between our calculations and those of FBC.  The first
is that our simulation incorporates radiative cooling, which changes
the temperature structure of collapsed regions and allows some of
the gas to condense into galaxies, removing it from the IGM.
The second is that our analysis incorporates photoionization,
while FBC only include collisional ionization.
As we can see from Figures~\ref{CNOfrac}-\ref{BFvsMi}, photoionization
makes an important difference for oxygen, nitrogen, and carbon,
though it is less significant for higher ionization species.
FBC also assume a much higher metallicity, with an
oxygen abundance of 0.5 solar, but this difference just shifts
absorber column densities by a factor of five.
Our input physics is similar to that of HGM,
who incorporate radiative cooling and photoionization using the
BF92 X-ray background but do not include a UV background.

\subsection{Calculating absorption}

The transmitted X-ray flux is
\begin{equation}
F(\nu_0) = F_c(\nu_0) \left(1-e^{-\tau(\nu_0)}\right),
\label{eqn:fnu}
\end{equation}
where $F_c$ is the quasar's X-ray continuum multiplied by the
instrument response function and $\tau(\nu_0)$ is the optical
depth at observed frequency $\nu_0$.
We will assume that the continuum can be determined and divided out,
so we define $F_c(\nu_0)\equiv 1$ in equation~(\ref{eqn:fnu}) and
refer to the transmitted flux 
$F(\nu_0) = \left(1-e^{-\tau(\nu_0)}\right)$
as the simulated spectrum.
Strong atomic features are rare in this wavelength regime, so
continuum estimation is relatively straightforward if the
signal-to-noise ratio of the spectrum is high enough.

The absorption spectrum is determined by
a convolution of the ion number density along the
line of sight with the absorption line profile,
\begin{equation}
\tau(\nu_0) =
\sum_{Z,I,l} \int \dd x \frac{\pi e^2}{m_e c} f_{l} \phi(x, \nu_0)
n_{Z,I}(x),
\label{eqn:spectrum}
\end{equation}
where $Z, I, l$ label the atomic number, the ionization stage, and
the transition line of the ion, respectively, $f_l$ is the oscillator
strength of the transition, and $n_{Z,I}(x)$ is the
ion number density. 
If we assume that the relative
abundances of heavy elements are similar to those of
the Sun, we can write
\begin{equation}
n_{Z,I}(x)=n_{H}(x) Y_{Z\odot} {\cal F}_{Z,I}(x)
                               \left(\frac{Z}{Z_\odot}\right) ~,
\end{equation}
where ${\cal F}_{Z,I}(x)$ is the ionization fraction for
the $(Z,I)$ ion computed from the density and temperature at position $x$,
$n_{H}(x)$ is the number density of hydrogen atoms,
$Y_{Z\odot}$ is the solar abundance of element $Z$ by number relative
to hydrogen, and $Z/Z_\odot$ is the metallicity relative to solar.
(We regret the use of $Z$ for two different purposes in the same
equation but see no better alternative).
The line profile $\phi(x, \nu_0)$ is
\begin{equation}
\phi(x,\nu_0)=\varphi\left[T(x),
\left(1+z(x)+\frac{u(x)}{c}\right)\nu_0\right],
\end{equation}
where $z(x)$ is the Hubble redshift at $x$, $u(x)$ is the peculiar velocity,
and
\begin{equation}
\varphi_{l}[T,\nu]=
\frac{e^{-(\nu-\nu_{l})^2/\Delta\nu_D^2}}{\sqrt{\pi}\Delta\nu_D}
\end{equation}
is the thermal broadening profile with
Doppler linewidth $\Delta\nu_D = \nu_{l} \sqrt{2kT/m c^2}$,
where $m$ is the ion mass.
We do not incorporate damping wings in the line profile, but they
are negligible for these weak absorption lines.

As described in \S 2.1, the average density, temperature, and
velocity of gas for each of the 5000 bins in a line of sight were
extracted from the simulation using TIPSY. For each ion of interest,
we calculate the ionization fraction for these temperature and density values
by interpolating among the Cloudy outputs. The optical depth along each
line of sight is then obtained by carrying out the convolution,
equation~(\ref{eqn:spectrum}), numerically.  We consider each
ionic species in isolation, so in practice we do not perform the
sum in equation~(\ref{eqn:spectrum}), only the integral.
Because X-ray forest lines are rare and the bright quasars amenable
to study with present instruments are at low redshifts, there will
usually not be much ambiguity in identification of features.

\subsection{Absorbers}

\begin{figure}
\centerline{
\epsfxsize=5.5truein
\epsfbox{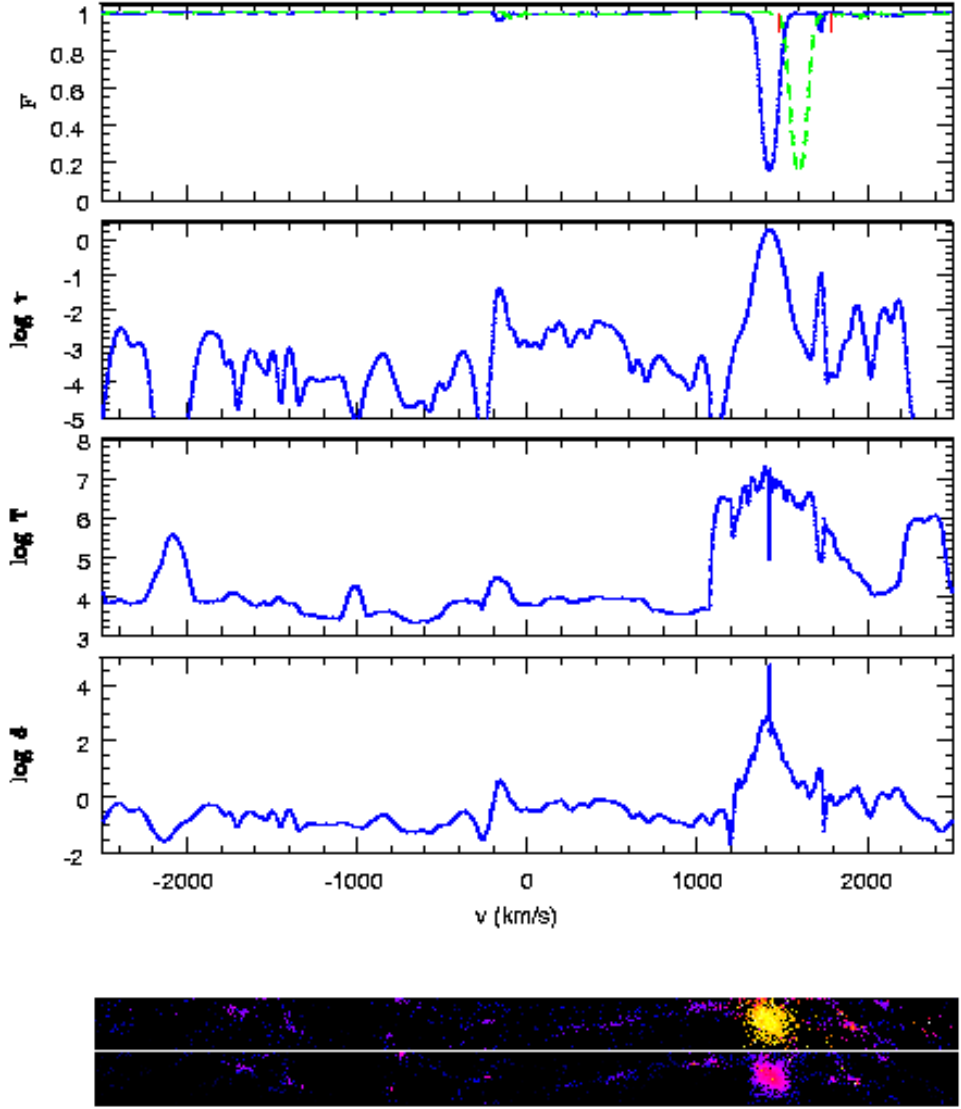}
}
\caption{
A Simulated spectrum for a line of sight with a strong \ovii\ absorber. From
top to bottom the panels show flux,
optical depth, temperature and overdensity along the line of sight,
as a function of spatial position expressed in velocity coordinates.
The two strips at the bottom show the gas particles from the simulation in a
rectangular prism $1\hmpc \times 1\hmpc \times 50\hmpc$ centered on this line
of sight, color coded by temperature (above) and density (below) (the
color scheme is the same as that of Fig.~\ref{zoombox}). 
Solid curves in all panels are calculated in real space
(no peculiar velocities) to allow the unambiguous identification of
features from panel to panel.  In the flux panel, the dashed line
shows the redshift space spectrum, and ticks mark the region
identified by our threshold algorithm as an ``absorber'' in
the redshift space spectrum.
}
\label{simspec}
\end{figure}

Figure~\ref{simspec} shows a sample \ovii\ absorption spectrum, one
that contains a strong feature.  As the lower panels demonstrate, this
absorption feature arises in the hot gas halo of a moderate sized
group, with a typical gas temperature of a few$\times 10^6\K$.
For clarity, we have ignored peculiar velocities when calculating these
profiles,
but we include thermal broadening in the optical depth and flux profiles.
The dashed line in the top panel
shows the redshift-space spectrum, in which 
the absorption feature is shifted by peculiar
velocity but not changed substantially in width or depth.
In all of our statistical calculations below, we identify absorbers
and measure their properties from the redshift-space spectrum,
including peculiar velocities.

We identify absorbers using a simple threshold algorithm:
an absorber is defined as a region where all
the pixels have absorption above a given threshold.
Because X-ray forest absorbers are likely to be unresolved
or nearly unresolved, at least for the foreseeable future,
more elaborate algorithms that deblend distinct peaks or
fit superpositions of Voigt profiles are unnecessary for our
present purposes.  We adopt an absorption threshold of 0.01,
hence absorbers consist of contiguous pixels with transmitted
flux less than 0.99.
We have checked the effect of using different thresholds and found
that while the number of weak absorbers is slightly suppressed
by a higher threshold, the number of strong absorbers is insensitive
to the threshold value.
In Figure~\ref{simspec}, the
boundary of the absorber is marked by ticks on the spectrum.

The equivalent width of an absorber is
\begin{equation}
W=\int_{v_{\rm min}}^{v_{\rm max}} \dd v 
\left(1-e^{-\tau(v)}\right),
\end{equation}
where $v=c z + u$, and $\tau$ is the optical depth. The equivalent
width defined in this way has velocity units.
Conversions to other useful units are
$W=6.67(W/100\kms)(\lambda/20\hbox{\AA})\mang = 0.33(W/100\kms)(E/1\keV) \ev$,
where $\lambda$ and $E$ are the observed line wavelength and energy,
respectively.  
For optically thin absorbers, there is a linear
relation between equivalent width and column density,
\begin{equation}
\label{col-width}
W_{\rm  thin}= \frac{\pi e^2}{m_e c} \lambda f_{Z,I} N_{Z,I} ~,
\end{equation}
where $N_{Z,I}$ is the column density and $f_{Z,I}$ is the
transition oscillator strength.
Since X-ray forest lines are weak and fairly broad, equation~(\ref{col-width})
is usually a good approximation.
However, the equivalent widths of the strongest absorbers are depressed
by saturation.  Figure~\ref{colO78} demonstrates this effect, plotting
the ratio of actual equivalent width to the optically thin value
of equation~(\ref{col-width}) for the \ovii\ and \oviii\ 
absorbers along our 1200 lines of sight.  The upper panels show results using
our standard metallicity $Z=0.1Z_\odot$, and the lower panels show results
using $Z=Z_\odot$, i.e., a factor of ten increase in absorber column
densities.  For \ovii, saturation becomes significant above
$N \sim 2\times 10^{15}\;\cm^{-2}$, $W \sim 80\;\kms$, and by
$N \sim 4\times 10^{15}\;\cm^{-2}$ the equivalent width is typically
depressed by a factor of two.  There are very few absorbers in this 
regime for $Z=0.1 Z_\odot$, but a significant number for $Z=Z_\odot$.
Saturation of \oviii\ lines does not become significant until 
$N\sim 10^{16}\;\cm^{-2}$, and we find few absorbers of this
column density even with $Z=Z_\odot$.
We will generally present our results in terms of equivalent width
(including saturation effects)
because this quantity is more directly determined from observations,
but we also label our plots by the corresponding optically thin column
density of equation~(\ref{col-width}), to aid with physical interpretations
and to compare with other numerical and analytic predictions.

\section{Equivalent Width Distributions and the Detectability
         of the X-ray Forest}
\label{sec:predictions}

\subsection{Equivalent width distributions}
\label{sec:ewdist}

Figure~\ref{optdistO} presents our principal quantitative prediction,
the cumulative distribution of \ovii\ and \oviii\ absorbers as a
function of equivalent width.  We plot $dN/d\ln (1+z)$, which
is equivalent to the number of lines per unit redshift at $z=0$.
Adopting the BF92 background in place of the Mi98 one, or omitting the
additional UV background at longer wavelengths, makes a nearly
imperceptible difference to these curves.
As mentioned earlier, the number of weak absorbers would be lower
if we adopted a higher threshold for the line identification algorithm,
but the number of strong absorbers would be unaffected.
We also find that the peculiar velocities have little effect on the
equivalent width distribution, though they change the equivalent
widths of individual absorbers to some extent.
The solid lines in each panel show our standard case of uniform metallicity
with $Z=0.1Z_\odot$.  The shapes of the \ovii\ and \oviii\ distributions
are similar, with somewhat fewer high equivalent width lines for \oviii.
The number of \ovii\ lines drops to one per
unit redshift at $W \sim 40\kms$ (25$\kms$ for \oviii) and
falls rapidly for larger $W$.  The paucity of stronger absorbers
is not surprising: for $Z=0.1Z_\odot$, a 1 Mpc path length through
a medium with $n_H=10^{-5}\cm^{-3}$ ($\delta_b=60$) and an \ovii\
fraction of 0.5 yields a column density of 
$1.1\times 10^{15}\cm^{-2}$,
or $W \sim 40\kms$, and longer path lengths at such high density and \ovii\
fraction are rare.

Dotted curves in Figure~\ref{optdistO} show predictions for a solar
metallicity IGM.  The increase in oxygen abundance
boosts the column density of each absorber by the same factor,
and at low $W$ the equivalent width distribution simply shifts to the
right by a factor of ten.  At high $W$ the rightward shift is smaller,
since saturation begins to suppress the linear relation between
equivalent width and column density; the high-$W$ tails of the solar
metallicity distribution functions are therefore steeper.
Because the equivalent width
distribution falls steeply at high $W$ even for $Z=0.1Z_\odot$, 
a modest increase in metallicity
can still translate into a large increase in the predicted number
of systems above some threshold --- i.e., a small horizontal shift
creates a large vertical shift.  Labeled vertical error bars in
Figure~\ref{optdistO} show representative $5\sigma$ detection thresholds
for several different X-ray satellites, which we discuss in
\S\ref{sec:capabilities} below.  Unfortunately, one can already see 
that a comprehensive
study of the X-ray forest with current instruments will be challenging
unless the IGM has near solar metallicity, which seems unreasonably
optimistic.  The remaining curves in Figure~\ref{optdistO} are discussed
at the end of the Section, where we consider the potential impact
of non-uniform metallicity.

The equivalent width distributions for the other ions listed
in Table~\ref{ions} are shown in Figures~\ref{optdistCN}
and~\ref{optdistNeSiFe}, computed assuming 0.1 solar metallicity
and the Mi98+UV background.
The number of high equivalent width absorbers is generally much
smaller than for \ovii\ or \oviii.  Roughly speaking, one can think
of the lower element abundances and, in some cases, weaker transition
oscillator strengths as shifting the previous equivalent width distributions
towards lower $W$, though of course the distribution also depends on
the occurrence of density and temperature regimes where these ions
have a high fractional abundance.
As a guide to relative observability, we list in Table~\ref{ions}
the equivalent width threshold above which we find one absorber
per unit redshift in our simulated spectra for our standard assumptions.
This threshold is roughly proportional to the adopted metallicity, though
it would increase more slowly above $W \sim 100\kms$ because of
saturation (see Figure~\ref{colO78}).
\ion{C}{5} and \ion{C}{6} have the strongest absorption after oxygen,
but their strong resonance lines lie at 40.3\AA\ and 33.7\AA, respectively,
which restricts their observability to relatively low redshift quasars
and puts them in a range of poorer instrument sensitivity even at zero
redshift.  \ion{Fe}{17} and \ion{Ne}{9} produce the next strongest 
absorption; their transition energies are higher, but the predicted
equivalent widths are a factor $\sim 2$ lower than those of \ion{C}{5}
and \ion{C}{6}.

While the focus of this paper is X-ray absorption, the (1032\AA, 1038\AA)
UV doublet of \ovi\ is another potential tracer of intergalactic gas
at these densities and temperatures, and is observable from
Hubble Space Telescope or the Far Ultraviolet Spectroscopic
Explorer (FUSE).  Recent observations have detected a number of
\ovi\ absorbers towards low redshift quasars, implying a high
number density $dN/dz\sim 20-50$ of lines with rest-frame equivalent
width $W_r > 30\mang$ \citep{Oeg00,TSJ00,TS00,T01,S02}.
Figure~\ref{O6} shows the cumulative equivalent width distribution
of the 1032\AA\ line of \ovi\ absorbers predicted by our simulation,
together with
data points from \cite{TSJ00} and \cite{S02} (taken from figure 2
of \citealt{Cen01}).  
We have used the Mi98+UV background, but the results are not very
sensitive to this choice.  As in Figure~\ref{optdistO}, solid and
dotted curves show predictions for uniform metallicity
$Z=0.1Z_\odot$ and $Z=Z_\odot$, respectively; long-dashed
and dot-short dashed curves show predictions that incorporate non-uniform
metallicity with typical $Z\sim 0.1Z_\odot$, 
as discussed at the end of the Section.
For any of the metallicity models except $Z=Z_\odot$, the simulation
results are in good agreement with the observations, though
we caution that our threshold algorithm may not
be as good a match to the analyses of these higher resolution, UV data.
The dot-long dashed curve of Figure~\ref{O6} shows the results of \cite{fang01},
who carried out similar calculations using a different numerical 
method, with a metallicity model similar to that shown by our
dot-short dashed line.
The agreement between the two calculations is remarkably good.
These results also agree fairly well with those of \cite{Cen01}, though
Cen et al.\ predict somewhat less absorption than observed.
Simultaneous detections of \ovi, \ovii, and \oviii\
could provide insight into the physical state of the
absorbing medium; we discuss the expected correlations between
these ionization phases in \S\ref{sec:physics} below.

In principle, \ovi\ absorption can also be studied with X-ray
telescopes, using the KLL resonance of the photoionization cross-section,
i.e., excitation-autoionization via the 1s $\to$ 2p transition
resulting in 1s2s2p (KLL) resonance.\footnote{This interesting possibility
was pointed out to us by Jordi Miralda-Escud\'{e} and Anil Pradhan.}
The strongest transition has $\lambda = 22.05$ \AA, $f=0.408$
\citep{Pradhan00}.  The dashed line at the lower left of
Figure~\ref{O6} shows the
equivalent width distribution for this absorption.
It is weaker than the corresponding \ovii\ and \oviii\ absorption
(compare to Figure~\ref{optdistO}),
primarily because of the lower fractional abundance of \ovi.
X-ray detection of intergalactic \ovi\ absorption is therefore
unlikely, at least without a substantial increase in
instrumental sensitivity.

The closest point of comparison for our calculation is that of HGM,
who used a different numerical method but similar input physics.
They present their \ovii\ and \oviii\ predictions in the form of
differential distributions, whereas we have shown cumulative distributions
in Figure~\ref{optdistO},
but we have also created differential plots (not shown) and compared
results.  At small equivalent width, we have excellent agreement with HGM.
However, the number of strong absorbers in our simulation 
(for our standard metallicity assumption $Z=0.1Z_\odot$) falls short
of the number in theirs by a factor of a few --- or, equivalently, for
absorbers that have a frequency $\sim 1$ per unit redshift, we predict a lower
equivalent width by a factor $\sim 2-4$.  
The cosmological parameters adopted by HGM are slightly different from
ours, in the direction of increasing the
number of absorbers (they have $\sigma_8=0.97, h=0.7, \Omega_b=0.05$).
Our box size of $50\hmpc$ is intermediate between those of the
two simulations used by HGM, $64\hmpc$ and $32\hmpc$, but close
enough to their larger box that this difference seems unlikely to
have a large systematic effect.
The total number of absorbers in the simulations at the equivalent
widths where we disagree is small, and it could be that the difference
simply reflects statistical fluctuations in the structure present
in the two simulation volumes.
However, we think that the most likely source of difference is the treatment
of IGM metallicity: we assume a uniform metallicity of 0.1 solar,
whereas HGM track the spread of metals in their simulation, so
that the metallicity varies with spatial position.
They comment that typical metallicities are $\sim 0.1 Z_\odot$ in the
density and temperature range that dominate X-ray forest absorption,
but they do not plot the scatter in metallicity, and
it could be that the strongest absorption occurs in regions where
the metallicity is higher than average.
\cite{CO99b} find a substantial scatter in metallicity at fixed
overdensity in their hydrodynamic simulations, with a mean of
$0.1Z_\odot$ at $\delta \approx 20$ and $1\sigma$ variations
of a factor of 2--3.
This effect would rather naturally explain the difference between
our results and HGM's, with agreement at low $W$ but
disagreement in the steeply falling tail of the distribution.

To address this point quantitatively, we show two additional curves in
each panel of Figure~\ref{optdistO}, 
calculated for a non-uniform metallicity with statistical
properties based on figure~2 of \cite{CO99b}.  Dashed curves show the
impact of metallicity scatter alone.  We compute them by drawing 
the metallicity of each line of sight independently from a log-normal
distribution with $\langle {\rm log}\;Z/Z_\odot\rangle = -1$ and
$\sigma_{{\rm log} z}=0.4$.  Because we assign a single metallicity
to a given line of sight, the fluctuation above or below the mean is
coherent across each absorber.  At low equivalent widths, where the 
distribution functions are shallow, metallicity scatter has little
impact.  For rare systems, however, the value of $W$ at a given value of
$dN/d{\rm ln}(1+z)$ increases by a factor of $2-3$, since
there are more intrinsically low column density systems to scatter
to high metallicity than vice versa.  Dot-dashed curves show the effect
of incorporating a density-dependent mean metallicity,
with $\langle {\rm log}\;Z/Z_\odot\rangle = -1.66 + 0.36\; {\rm log}\; 
\delta$,
while retaining log-normal scatter about the mean metallicity as before.
This change depresses the distribution function at lower $W$, where
systems typically arise in lower density gas.  We conclude that the
differences in metallicity treatment could plausibly account for most
of the difference between our calculations and that of HGM, and
that a metallicity distribution similar to that of \cite{CO99b}
would increase the predicted equivalent widths of rare systems (at
fixed line density) by a factor $\sim 2$ relative to our uniform,
$Z=0.1Z_\odot$ case.  
We also show \ovi\ predictions for these non-uniform metallicity models
in Figure~\ref{O6}; the qualitative effect of non-uniform metallicity 
is similar, but the quantitative impact at the equivalent width
thresholds of the observational data points is modest because these
points do not lie on the extreme tail of the distribution.

FBC predict a much higher occurrence of strong \ovii\ and \oviii\
absorption than either
HGM or our standard model, but this difference is mainly attributable
to their higher assumed oxygen abundance of $0.5Z_\odot$,
also adopted in the analytic calculations of FC and 
\cite{PL98}.  (These papers also use an older solar abundance for 
oxygen that is slightly higher than the one adopted here. In our units,
their abundance would be $0.576 Z_\odot$.)
Direct comparison of our Figure~\ref{optdistO} to FBC's figure 13 is
difficult because they show differential distributions of column density
while we show cumulative distributions of equivalent width (including
saturation effects).  However, T.\ Fang has kindly provided the
FBC \ovii\ data in numerical form, and we have made direct comparisons
of the column density distributions.
At high column densities ($N \ga 5\times 10^{15}\;\cm^{-2}$), the
FBC results for oxygen abundance $0.576Z_\odot$ lie very close 
to our results for $Z=Z_\odot$.
(Recall that in this regime, the equivalent widths in Figure~\ref{optdistO}
have typically been depressed by a factor of two or more relative
to their optically thin values.)
The fact that the predictions match for metallicities that differ by 
a factor of 1.7 can plausibly be attributed to the presence in our
simulation of radiative cooling, which
converts some of the hot gas in collapsed halos into cold gas and stars.
At lower column densities, our solar metallicity model predicts substantially
more absorbers than FBC 
(e.g., by a factor $\sim 7$ at $N\sim 10^{14}\;\cm^{-2}$).
The difference in the shapes of the distribution functions can 
be plausibly attributed to the impact of photoionization,
which becomes important at the lower densities and temperatures
characteristic of weaker \ovii\ systems.

Once we allow for differences in assumptions about metallicity,
we find the general level of agreement between our results and those
of HGM and FBC (and \citealt{Cen01} and \citealt{fang01} for \ovi)
to be encouraging.
It suggests that, at the factor of two level,
the X-ray forest predictions from hydrodynamic simulations are not 
sensitive to the numerical methodology, or even to the inclusion 
of radiative cooling and star formation, and that photoionization
is significant for weaker absorption systems but not for the strongest
absorbers.  The largest uncertainty in the predictions is 
the poorly known metallicity distribution of the IGM, and for \ovi,
the strength of the metagalactic flux at $\sim$ 9 Ryd.
FBC compare their numerical results to the analytic, Press-Schechter 
based approach of FC and \cite{PL98} in some detail.  They find that the 
analytic approach works well for high-excitation ions
(e.g., \ion{Fe}{25} and \ion{Fe}{26}) and for the highest
column density systems of lower excitation ions
(e.g., $N\ga 10^{16}\;\cm^{-2}$ for \ovii\ and \oviii),
while it tends to underestimate the occurrence of weaker absorption,
which often arises outside of the collapsed halos modeled by
the Press-Schechter formalism.

\subsection{Observational capabilities}
\label{sec:capabilities}

The best spectral resolution of the
current and next generation X-ray telescopes is
$R\equiv \lambda/\Delta\lambda \sim 1000$, or about
$300 \kms$ in velocity units.  Most absorbers, therefore, are likely to
be unresolved.
In a quasar spectrum, the minimum detectable equivalent width for an
unresolved absorption feature is
\begin{equation}
W_{\mathrm{min}}= \left(\frac{S}{N}\right)_{\rm min}
\left(\frac{E}{R A F_x t}\right)^{1/2},
\label{eqn:wmin}
\end{equation}
where $\left(\frac{S}{N}\right)_{\rm min}$ is the minimum signal-to-noise
ratio required for detection,
$E$ is the observed energy of the
line, $t$ is the integration time, $R$ is the spectral resolution,
$A$ is the effective area, and
$F_x$ is the photon flux, e.g., in units of
photons $\cm^{-2} \Sec^{-1} \keV^{-1}$.
Equation~(\ref{eqn:wmin}) yields an equivalent width in the same
energy units used for $E$ and $F_x$.  
An X-ray spectrum of a moderate redshift quasar might have
$N_{\rm res} \sim 500$ resolution 
elements in which absorption could
potentially fall, so the $S/N$ threshold for a ``blind'' detection
must be fairly high to avoid the spurious identification of
noise peaks as absorption systems.
We therefore adopt $S/N=5$ as a threshold for secure detection in 
our discussion below.
A $4\sigma$ feature would have a probability 
$\sim 3\times 10^{-5} N_{\rm res}$ of arising by chance,
assuming Gaussian statistics, so it would be sufficient for a
tentative detection.

The characteristics of each X-ray telescope depend on the choice of
instrument setting (grating, detector, etc.).
We will not discuss these issues in detail here, but we list in
Table~\ref{instrument} some representative value of effective area and
resolving power for {\it Chandra}\footnote{http://chandra.harvard.edu},
{\it XMM-Newton}\footnote{http://xmm.vilspa.esa.es},
{\it Constellation-X}\footnote{http://constellation.gsfc.nasa.gov}
and {\it XEUS}.\footnote{http://astro.estec.esa.nl/SA-general/Projects/XEUS}
To assess detectability thresholds, we assume a quasar flux
\begin{equation}
F_x= F_0 \left(\frac{E}{E_{0}}\right)^{-\Gamma},
\end{equation}
and we adopt as an example the X-ray bright quasar
H1821+643 ($z=0.297$), with $F_0=2 \times
10^{-3}$ photons $\cm^{-2} \Sec^{-1} \keV^{-1}$ at $E_0=1\keV$,
and $\Gamma=2.35$ (S.\ Mathur, private communication; see 
\citealt{mathur03} for more detailed discussion of H1821+643).
We have not accounted for the effects of Galactic HI absorption,
which depresses the flux by $\sim 30\%$ at wavelengths relevant
to \ovii\ or \oviii\ detection.
Values of $W_{\rm min}$ for $S/N=5$ and an integration time of 500 ksec
are listed in Table \ref{equivwidth}, for a variety of observed wavelengths.

Figures~\ref{optdistO}--\ref{optdistNeSiFe} show that
for a given path length it is easiest to detect
\ion{O}{7} and \ion{O}{8} absorbers. However, as shown in
Table~\ref{equivwidth}, even a 500 ksec integration with the
brightness of H1821+643 yields a minimum detectable equivalent width
of $\sim 270\kms$ for {\it Chandra} and $\sim 160\kms$ for {\it XMM-Newton}.
These limits are marked by vertical bars in Figure~\ref{optdistO}.
Assuming a metallicity of 0.1 solar, the strongest
absorbers in our simulated spectra have $W \sim 200\kms$, which
are at best marginally detectable by {\it Chandra} and {\it XMM-Newton}.
The curves with metallicity scatter are arguably more realistic
than the constant metallicity case, for the reasons discussed
by \cite{CO99b}, and it is possible that enriched regions of
the IGM have enhanced ratios of $\alpha$-elements like oxygen
relative to iron.  Both of these changes would improve the 
prospects of detection.

A search for \ovii\ and \oviii\ absorption associated with known \ovi\
absorption (from UV observations) holds considerably more promise than
a blind search for X-ray forest absorption, for two reasons.
First, when searching at a redshift that is known a priori, the
required $S/N$ for a significant detection is lower, $\sim 2-3\sigma$
instead of $\sim 5\sigma$, and at these equivalent widths the number
of systems rises sharply with decreasing detection threshold.
Second, as we will show shortly, the strongest \ovi\ absorbers
frequently have strong associated \ovii\ absorption, though this
is less true for \oviii.

{\it Constellation-X}, which is likely to be the next major advance
in capability for X-ray spectroscopy, is now under design.
For the observing conditions assumed in Table~\ref{instrument},
the minimum detectable
width for {\it Constellation-X} is about 25 km/s; thus, for a metallicity
of 0.1 solar, we predict that Constellation-X could detect 
2-3 \ovii\ absorbers and $\sim 1$ \oviii\ absorber
per unit redshift along each random line of sight, at $5\sigma$ significance.
The numbers rise slightly if one adopts the \cite{CO99b} metallicity-density
relation, and if the metallicity
is as high as solar then tens of absorber should be detected per line of
sight.  The powerful {\it XEUS} X-ray telescope, with a
typical equivalent width threshold of $\sim 10\kms$,
should be able to find large numbers of absorbers.

\section{Physical Properties and Correlations of Oxygen Absorbers}
\label{sec:physics}

We now turn to the physical state of the gas producing oxygen absorption,
which appears to be the element most amenable to observational
study.  Figures~\ref{zoombox} and~\ref{boxmap} are a useful starting point.
Figure~\ref{zoombox} shows the gas particles in the simulation
color-coded by temperature (upper left), density (upper right),
\ovii\ ion fraction (lower left), and \oviii\ ion fraction (lower right).
We have zoomed in on a $(25\hmpc)^3$ sub-volume of the simulation
in order to show finer details of the structure.
Gas in virialized halos has been shock-heated to typical
temperatures of $\sim 10^6-10^7\K$, while gas in the filaments
connecting these halos is typically somewhat cooler, from $\sim 10^5\K$
to a few$\times 10^6\K$.  The regions of high \ovii\ and \oviii\ 
ionization fractions trace the general structure of these filaments
and halos.  The \ovii\ fraction shows fairly sharp peaks in the dense
inner regions of the halos, though there is by no means a one-to-one
correspondence between density peaks and peaks of \ovii\ fraction.
\oviii\ tends to dominate in the outer regions of halos and filaments,
where the density is lower.  The relatively massive, hot
group at the center of the upper panels has low \ovii\ and \oviii\ fractions
because its temperature
is high enough to fully ionize the oxygen to \oix.

Figure~\ref{boxmap} shows the projected column density of \ovii\
in the full $50\hmpc$ simulation cube, with threshold column
densities in the four panels of $10^{14}$, $10^{14.5}$,
$10^{15}$, and $10^{15.5}\cm^{-2}$.  The assumed IGM
metallicity is 0.1 solar.  At the $10^{14}\cm^{-2}$ threshold, \ovii\
traces a rich network of shock-heated filaments.
At $10^{14.5}\cm^{-2}$, the stronger filaments remain, but
the weaker ones disappear from view.
By $10^{15}\cm^{-2}$, the distribution consists mainly of distinct
clumps, occasionally connected by a more diffuse medium.
At $10^{15.5}\cm^{-2}$, only the dense regions in collapsed halos
are visible.  The redshift path length through the simulation
volume is only $\Delta z=0.0167$, so the covering factor for absorption
to a quasar of redshift $z$ would be larger by $z/\Delta z$, e.g.,
by a factor of 18 for $z=0.3$.
The column density thresholds in the last three panels
correspond roughly to the detectability thresholds for
{\it XEUS}, {\it Constellation-X}, and {\it Chandra}/{\it XMM-Newton},
given the assumptions discussed in \S\ref{sec:capabilities}.
If the IGM metallicity is $Z \sim 0.1Z_\odot$, therefore, current generation
X-ray telescopes have the potential to detect X-ray forest absorption
in high-density gas, {\it Constellation-X} can begin to reach into
intergalactic filaments, and {\it XEUS} can map the filamentary structure.
If the IGM metallicity is 0.3 solar, then the same panels represent
column density thresholds that are $\sim 0.5\;$dex higher, so
the {\it Chandra}/{\it XMM-Newton} threshold would correspond to
the lower left panel, the {\it Constellation-X} threshold to the
upper right, and the {\it XEUS} threshold to the complex network
of filaments in the upper left.

To quantify the physical properties of the oxygen absorption systems,
we calculate the average temperature and overdensity of the gas
associated with each absorber.
We first define a {\it pixel} temperature and
density, assigning to each pixel along the line of sight a
temperature and density that is an average of the gas density and
temperature contributing to that pixel weighted by optical depth,
\begin{eqnarray}
n_{\rm pix}(v) &=& [\tau(v)]^{-1} \int dx \left(\frac{d\tau}{dx}\right) n(x),\\
T_{\rm pix}(v) &=& [\tau(v)]^{-1} \int dx \left(\frac{d\tau}{dx}\right) T(x).
\end{eqnarray}
The absorber temperature and overdensity are then obtained
by averaging (with equal weight) over the pixels belonging to that absorber.
Thermal broadening allows gas of different densities and temperatures
to contribute to a given pixel, but the optical depth weighting ensures
that the average is not dominated by gas that is not actually responsible
for the corresponding absorption.

Figure~\ref{Ophysprpty} plots the temperature and overdensity of
each absorber against its equivalent width, for absorbers with $W \geq 10\kms$.
The assumed metallicity
is $Z=0.1 Z_\odot$; for a higher assumed metallicity one would,
in the optically thin approximation, simply shift each point horizontally
by the same factor by which the metallicity is increased.
Most \ovi\ absorbers have temperature $10^{4.2}\K < T < 10^{5.7}\K$,
\ovii\ absorbers have $10^{4.5}\K < T < 10^{6.5}\K$,
and \oviii\ absorbers $10^{5.2}\K < T < 10^{6.7}\K$.
For all three species, there is a correlation between equivalent
width and gas overdensity, though there are some notable outliers
from these correlations.
In the case of \ovii\ absorbers, the median overdensity
increases from $\delta \sim 10$ at $W \sim 10\kms$ to $\delta \sim 100$
at $W \sim 50\kms$, and the three absorbers with $W>100\kms$
all have $\delta > 100$.  The system that produces the strongest
\ovii\ and \oviii\ absorption has $\delta \sim 10^4$, but
this system is weak in \ovi\ because of its high temperature
($T \sim 2\times 10^6\K$).
In Figures~\ref{Ophysprpty}--\ref{O1-678}, special symbols mark the
five systems with the strongest \ovii\ absorption.  
The ``same'' absorber may have different density and temperature in 
different panels because of the optical depth weighting by the
corresponding ionic species and because the velocity ranges defined
in the different species may not be identical.

Figure~\ref{Ophysprpty} confirms the impression from
Figures~\ref{zoombox} and~\ref{boxmap} that strong \ovii\ and \oviii\
absorption arises in high density gas, more characteristic of
virialized halos than connecting filaments.\footnote{The value of
the ``virial'' overdensity is not very well defined, since different
papers choose the virial radius in different fashion, but most
definitions yield a local overdensity $\delta \sim 50-100$ at
the virial boundary.
For example, \cite{navarro97} find $\delta \sim 125$ for $\Lambda$CDM 
at the radius where the mean interior density is 200 
times the {\it critical} density and $\delta \sim 40$ at the radius where 
the mean interior density is 200 times the mean density.
(See their figure 3.
Their $\Lambda$CDM parameters, $\Omega_m=0.25$, $h=0.75$, and $\sigma_8=1.3$,
are significantly different from ours, in a direction that produces
denser halos.)}
Strong \ovi\ absorbers also tend to be fairly high overdensity,
though there are some strong \ovi\ systems with $\delta \sim 10-50$,
and the highest overdensity absorbers tend to be weak in \ovi\
because they are too hot.
For each species, the
strongest absorbers usually have temperatures fairly close
to the value where the corresponding ionic abundance peaks
in collisional equilibrium (see Figure~\ref{CNOfrac}), but
there is a spread in temperature at every value of $W$, and
for all three species there is a tail of weaker, lower temperature,
lower density absorbers for which photoionization plays an
important role.
Our conclusions about \ovi\ agree with those of \cite{Cen01}
and \cite{fang01}, in that collisional ionization dominates
for the strongest absorbers but photoionization becomes important
at slightly lower equivalent widths.
Our conclusions about \ovii\ and \oviii\ appear consistent
with those of HGM, to the extent that we can compare them.
The collisional equilibrium assumption used by FBC
in their simulation analysis should be fairly accurate for
the strongest \ovii\ and \oviii\ absorbers, but it breaks
down for the lower density absorbers associated with
filamentary structures.

With sufficiently good data, it should be possible to measure
absorption by multiple ionization stages of oxygen, or to obtain
limits on the relative absorption by different species.
Figure~\ref{Ocorr} shows the predicted correlations between \ovii\
and \oviii\ equivalent widths (top), \ovi\ and
\ovii\ (middle), and \ovi\ and \oviii\ (bottom).
There is some ambiguity in the definition of these correlations,
since the threshold algorithm will, in general, choose somewhat different
velocity ranges to mark the same absorber in different species, and
in some cases there may be an absorber above threshold in one species
but no corresponding absorber in the other.  In Figure~\ref{Ocorr},
crosses represent absorption in the velocity range defined by the
ion on the abscissa, and squares represent absorption in the velocity
range defined by the ion on the ordinate.  In many cases, the absorbers
defined in these two ways nearly coincide, so the square and cross
are close to each other, but in some cases, especially in the
\oviii\ vs.\ \ovi\ panel, there are absorbers in one species
with no obvious counterpart in the other.

The top panel shows a fairly good correlation between \ovii\ and \oviii\
absorption, as one might expect on the basis of
Figures~\ref{CNOfrac} and~\ref{BFvsMi}: there are substantial
ranges of temperature where both species have high fractional
abundance, and in this regime the density of both species
increases in proportion to the gas density.  However, the spread in IGM
temperatures does produce significant spread in the
\oviii/\ovii\ ratio.  In particular, there are some strong \ovii\
systems with relatively weak \oviii; these correspond to high density
gas where the temperature is too low to produce substantial \oviii\
by collisional ionization.

The \ovii\ vs. \ovi\ diagram shows a tight
core of points with a well defined correlation, tracing the underlying
correlation of both species with density (see Figure~\ref{Ophysprpty}).
The strongest \ovi\ absorbers tend to have strong associated \ovii,
which suggests that following up known \ovi\ absorbers is a promising
way to find \ovii\ X-ray forest absorption.  However, there are
some strong \ovi\ absorbers with weak \ovii, corresponding to dense,
relatively cool gas in which lower ionization species dominate.
Moreover, the strong \ovii\ absorbers ($W\ga 30\kms$) have a wide
range of \ovi\ strength, including quite low values, since the
\ovi\ fraction drops to nearly zero once the temperature exceeds
$5\times 10^5\K$.
Thus, while \ovi\ is a useful signpost for \ovii\ absorption,
it will not mark the positions of all or even most strong \ovii\
absorbers, nor should the existence of strong \ovi\ absorption
with weak \ovii\ come as a surprise.

Despite the correlations between \ovi\ and \ovii\ and
\ovii\ and \oviii, there is virtually no correlation between
\ovi\ and \oviii;
the variations
in temperature at fixed density that produce much of the
scatter in the top two panels drive the \oviii/\ovii\ and \ovi/\ovii\
ratios in opposite directions.
The lack of correlation in the bottom panel is not surprising
in light of Figure~\ref{CNOfrac} and~\ref{BFvsMi}, which
show only narrow ranges of temperature for which \ovi\ and \oviii\
both have substantial fractional abundance.

Figure~\ref{myhist} displays these predictions in a form that
is useful for interpreting X-ray forest searches at the redshifts of
known \ovi\ absorbers.  Upper panels show the distributions of \ovii\ 
and \oviii\ equivalent widths for systems with \ovi\ equivalent width
above $30\;\kms$ (103 m\AA).  Both distributions are bimodal, with
a main peak of cooler systems that have weak \ovii\ or \oviii\ and a 
secondary peak of hotter systems that have stronger X-ray
absorption.  Lower panels show the corresponding distributions for
an \ovi\ equivalent width threshold of $10\;\kms$ (34 m\AA).
Here the distributions are more dominated by weak \ovii\ and \oviii\ systems,
though there is still a tail of stronger absorbers.
Once again it is clear that following up strong \ovi\ absorbers is
a useful way of finding X-ray forest absorption, but 
that many \ovi\ absorbers, especially the weaker ones, have little 
associated \ovii\ or \oviii.

Figure~\ref{O1-678} shows the predicted correlations of
\ovi, \ovii, and \oviii\ equivalent width with \ion{H}{1} (Ly$\alpha$)
equivalent width, where in all cases we take the velocity ranges
defined by the \ion{H}{1} absorbers.  For \ovi\ there is a tight
correlation at low equivalent widths, reflecting the increase
of both \ion{H}{1} and \ovi\ with density in photoionized gas.
There is a spread of points towards higher \ovi\ associated
with shock-heated gas --- raising the temperature always decreases
the HI fraction at a given density, but it can increase the \ovi\
fraction.  At high equivalent widths the \ovi--\ion{H}{1} correlation flattens
and spreads, but it remains the case that most of the strong
\ovi\ systems have strong associated \ion{H}{1}, and vice versa.

There are hints of the same correlation present in the \ovii--\ion{H}{1}
diagram, but in this case the scatter is considerably larger, and
at the high equivalent width end (defined either by high \ion{H}{1} or by
high \ovii)
there is an enormous spread in \ovii/\ion{H}{1}, with essentially no
discernible correlation.  The \oviii-\ion{H}{1} diagram shows no
significant correlation or anti-correlation, except that the
strongest \ion{H}{1} systems all have \oviii\ absorption with $W>10\kms$.
Given the radically different temperature ranges required for \ion{H}{1}
and \oviii\ to exist in substantial abundances, it seems likely
that the \ion{H}{1} absorption is arising in the environs of the hotter
gas producing the \oviii\ absorption, rather than in the same gas.
Unfortunately, the large spread in \oviii/\ion{H}{1} and
\ovii/\ion{H}{1} suggests that it will be difficult to infer
the metallicity of X-ray forest systems directly from
line ratios.  The prospects are better
for inferring the metallicity in \ovi\ absorption systems, but these
trace only the cooler portions of the forest.

\section{Discussion}
\label{sec:discussion}

The advent of X-ray telescopes with large effective areas and instruments
for high resolution spectroscopy opens a new window on the intergalactic
medium, one that could reveal a large reservoir of previously hidden
baryons.  We have used a hydrodynamic simulation of the $\Lambda$CDM
cosmological model to predict the properties of the X-ray forest at $z=0$.
In agreement with previous studies based on numerical simulations
(HGM; FBC) and analytic methods \citep{PL98,FC00},
we find that the oxygen species \ovii\ and \oviii\ produce the strongest
X-ray forest lines, thanks to the high cosmic abundance of oxygen, the
oscillator strengths of the transitions at 21.60\AA\ (\ovii)
and 18.97\AA\ (\oviii), and the relatively frequent occurrence of
temperature and density regimes where these species have
large fractional abundance.
If the IGM metallicity at these temperatures and densities is $Z=0.1Z_\odot$,
as we assume in most of our calculations, then \ovii\ and \oviii\ are the
only X-ray forest lines likely to be detectable with existing instruments,
and even these will be difficult to study in any comprehensive fashion.
Other promising lines for study with future instruments include
the helium-like and hydrogen-like species of carbon, nitrogen, and neon
(\ion{C}{5}, \ion{C}{6}, \ion{N}{6}, \ion{N}{7}, \ion{Ne}{9}, \ion{Ne}{10}),
\ion{Fe}{17}, which has lower abundance but a large oscillator strength,
and possibly \ion{Si}{13}.  We have not examined highly ionized
species like \ion{Fe}{25} and \ion{Fe}{26}, which occur only at high
temperatures that are poorly represented in our $50\hmpc$ box
(see \citealt{FC00} for analytic predictions).

Our conclusions about the detectability of these X-ray forest lines
are summarized by the rightmost column of Table~\ref{ions},
which lists the equivalent width threshold at which we predict one
line per unit redshift, and Table~\ref{equivwidth}, which gives approximate
$5\sigma$ detection thresholds for observations with {\it Chandra},
{\it XMM-Newton}, {\it Constellation-X}, and {\it XEUS},  assuming
500 ksec exposures and the source spectrum of the X-ray bright
quasar H1821+643 ($z=0.297$).  
Target-of-opportunity searches that observe blazars in very
high states could potentially reach lower equivalent width thresholds
(Nicastro, private communication).
The full distribution functions for
absorber equivalent widths are shown in
Figures~\ref{optdistO}--\ref{optdistNeSiFe}.
The largest uncertainty in the predictions is the unknown metallicity
distribution of the IGM at the relevant temperatures and densities.
This distribution can itself be predicted by tracking the dispersal
of metals in cosmological simulations
(as done by, e.g., HGM and \citealt{Cen01}), but the uncertainties
in the escape of metals from the galaxies themselves make these
results quite uncertain.  We predict a lower frequency of detectable
X-ray forest lines than FBC, mainly because we assume
$Z=0.1Z_\odot$ instead of the higher metallicity found in the central
regions of hot X-ray clusters.  For the same metallicity, the two 
simulations give similar predictions for the number of strong \ovii\
absorbers, though the inclusion of radiative cooling and photoionization in 
our calculations tends to reduce the column densities of the strongest
systems and substantially increase the number of weak absorbers.
Our predicted \ovii\ equivalent width
distribution agrees with that of HGM at low $W$ but falls below it
at high $W$, perhaps because of differences in the assumed
cosmological parameters
and in the simulation methods, but perhaps because their strongest absorbers
arise in regions of higher than average metallicity.

\begin{figure}
\centerline{
\epsfxsize=5.5truein
\epsfbox{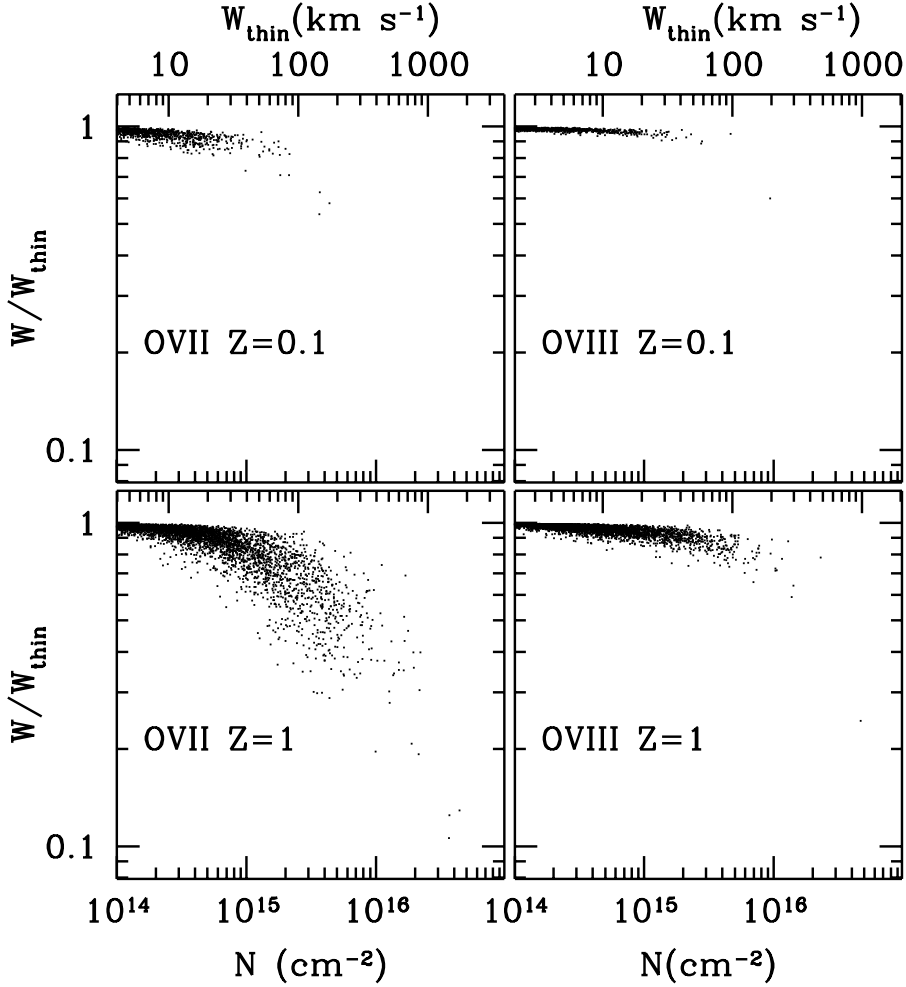}
}
\caption{The ratio of the equivalent width of absorbers 
to the optically thin equivalent width, as a function of column density,
for the \ion{O}{7} (left) and \ion{O}{8} (right) absorbers
along 1200 lines of sight through our simulation volume.
The top panels are based on our standard metallicity assumption
$Z=0.1Z_\odot$, while the bottom panels assume a solar metallicity
and are thus a factor of ten higher in column density.
We plot $W/W_{\rm thin}$, where 
$W_{\rm thin}= 3.985 (N/10^{14} \cm^{-2}) \kms $ for \ion{O}{7}
and $W_{\rm thin}=2.092 (N/10^{14} \cm^{-2}) \kms  $ for \ion{O}{8}.
The upper axis label shows the value of $W_{\rm thin}$ corresponding
to each column density.
}
\label{colO78}
\end{figure}

\begin{figure}
\begin{center}
\begin{minipage}{0.43\textwidth}
\epsfig{file=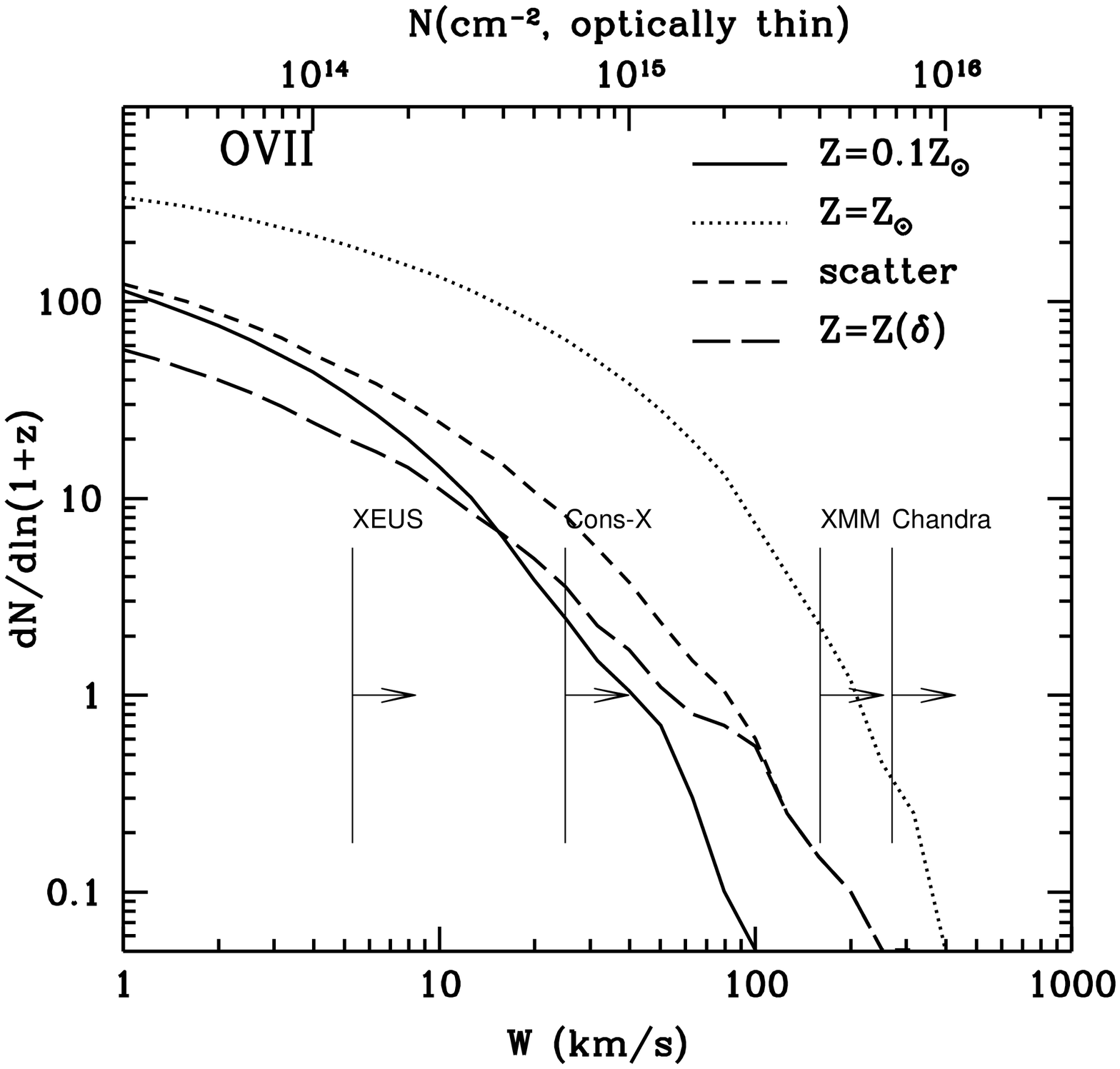,width=2.9in}
\end{minipage}
\begin{minipage}{0.43\textwidth}
\epsfig{file=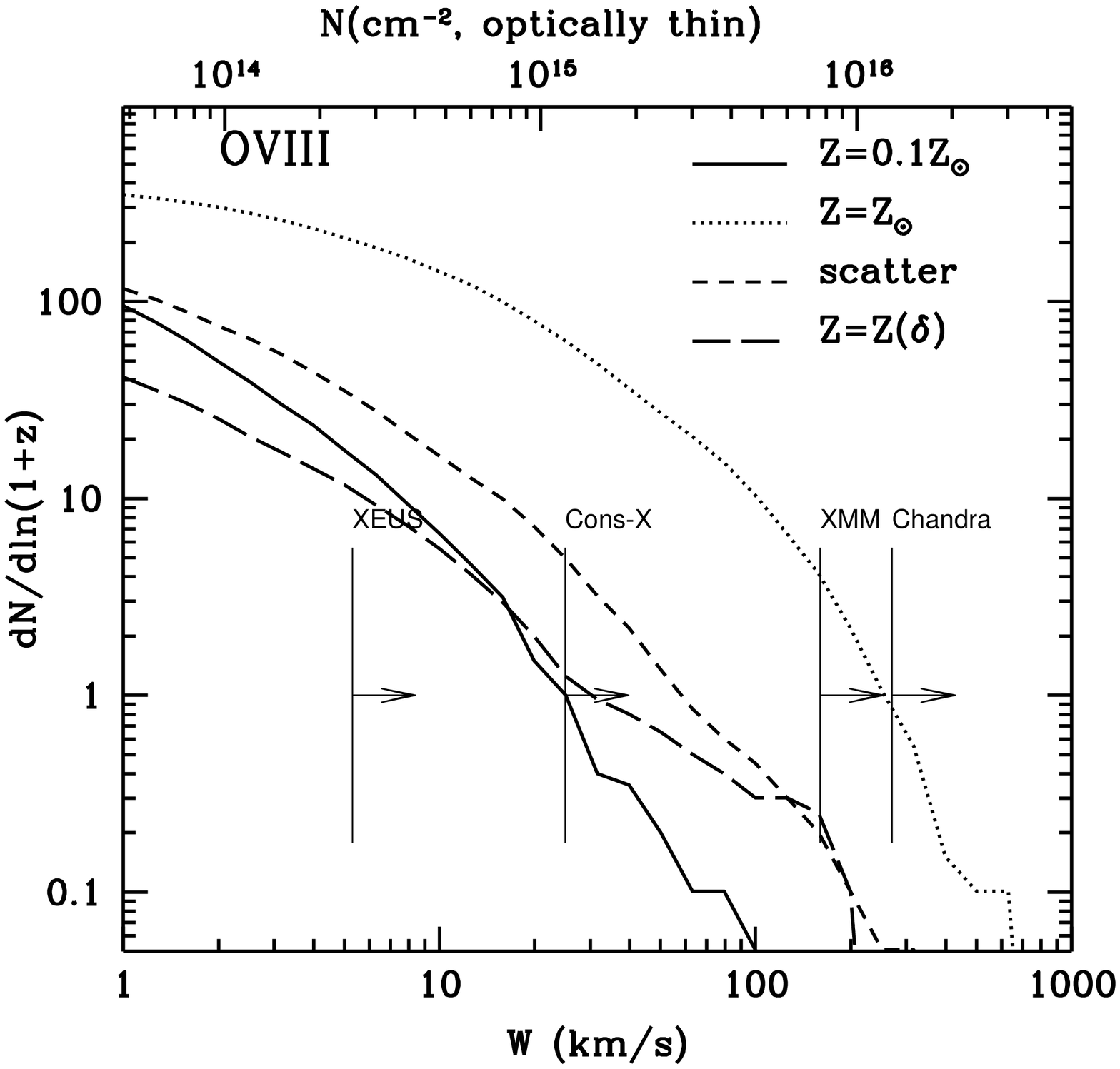,width=2.9in}
\end{minipage}
\end{center}
\caption{
The cumulative distribution function of \ion{O}{7} (left)
and \ion{O}{8} (right) equivalent widths,
i.e., the number of lines per unit redshift with equivalent
width greater than $W$, computed at $z=0$.
On the top of each panel, we mark column densities
corresponding to the given $W$ values for an optically thin absorber
(eq.~\ref{col-width}); above $N\sim 2\times 10^{15}\cm^{-2}$ these
are lower than the true column densities (see Fig.~\ref{colO78}).
All curves are calculated for the Mi98+UV background, but they
are insensitive to this choice.  Solid curves show our standard
case, uniform metallicity $Z=0.1Z_\odot$.  Dotted curves show
uniform metallicity $Z=Z_\odot$.
Short-dashed curves show the effect of scatter in metallicity; the
metallicity of each line of sight is drawn randomly from a
log-normal distribution with $\langle {\rm log}\; Z/Z_\odot \rangle = -1$,
$\sigma_{{\rm log} Z} = 0.4$.
Long-dashed curves show a calculation with the same metallicity scatter and
a trend of mean metallicity with gas overdensity ---
$\langle {\rm log} Z/Z_\odot \rangle = -1.66 + 0.36\; {\rm log}\; \delta$ ---
as advocated by \cite{CO99b}.
Vertical bars mark representative $5\sigma$ detection thresholds
for different X-ray satellites (see discussion in \S\ref{sec:capabilities}).
}
\label{optdistO}
\end{figure}

\begin{figure}
\centerline{
\epsfxsize=4.5truein
\epsfbox{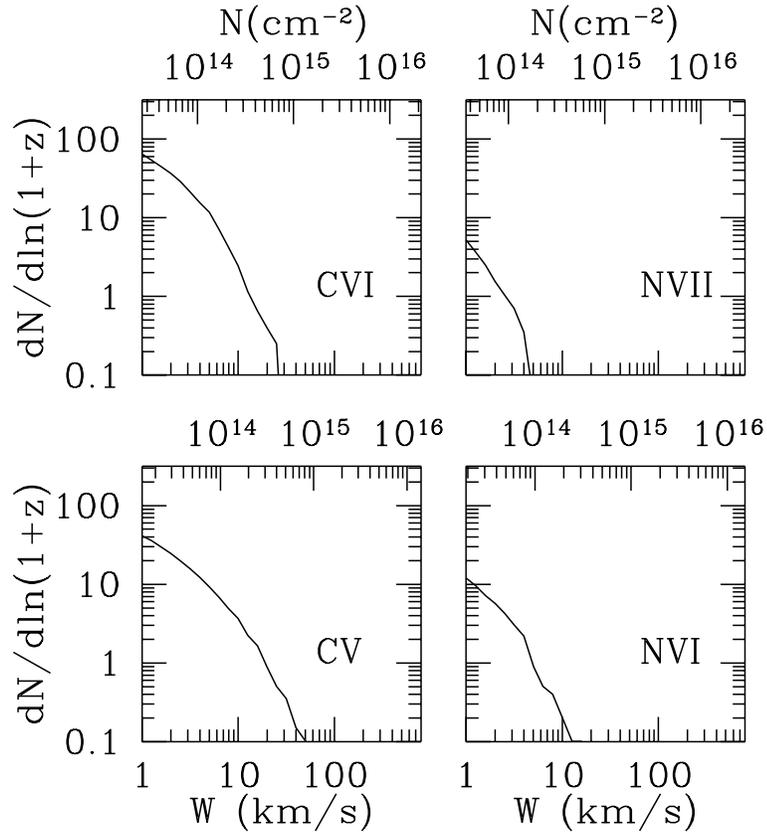}
}
\caption{
Cumulative equivalent width distributions of \ion{C}{5},
\ion{C}{6}, \ion{N}{6} and \ion{N}{7}, computed assuming
the Mi98+UV radiation background and $Z=0.1Z_\odot$.
The upper axis label shows the column density corresponding
to the given equivalent width for an optically thin absorber.
}
\label{optdistCN}
\end{figure}

\begin{figure}
\centerline{
\epsfxsize=4.5truein
\epsfbox{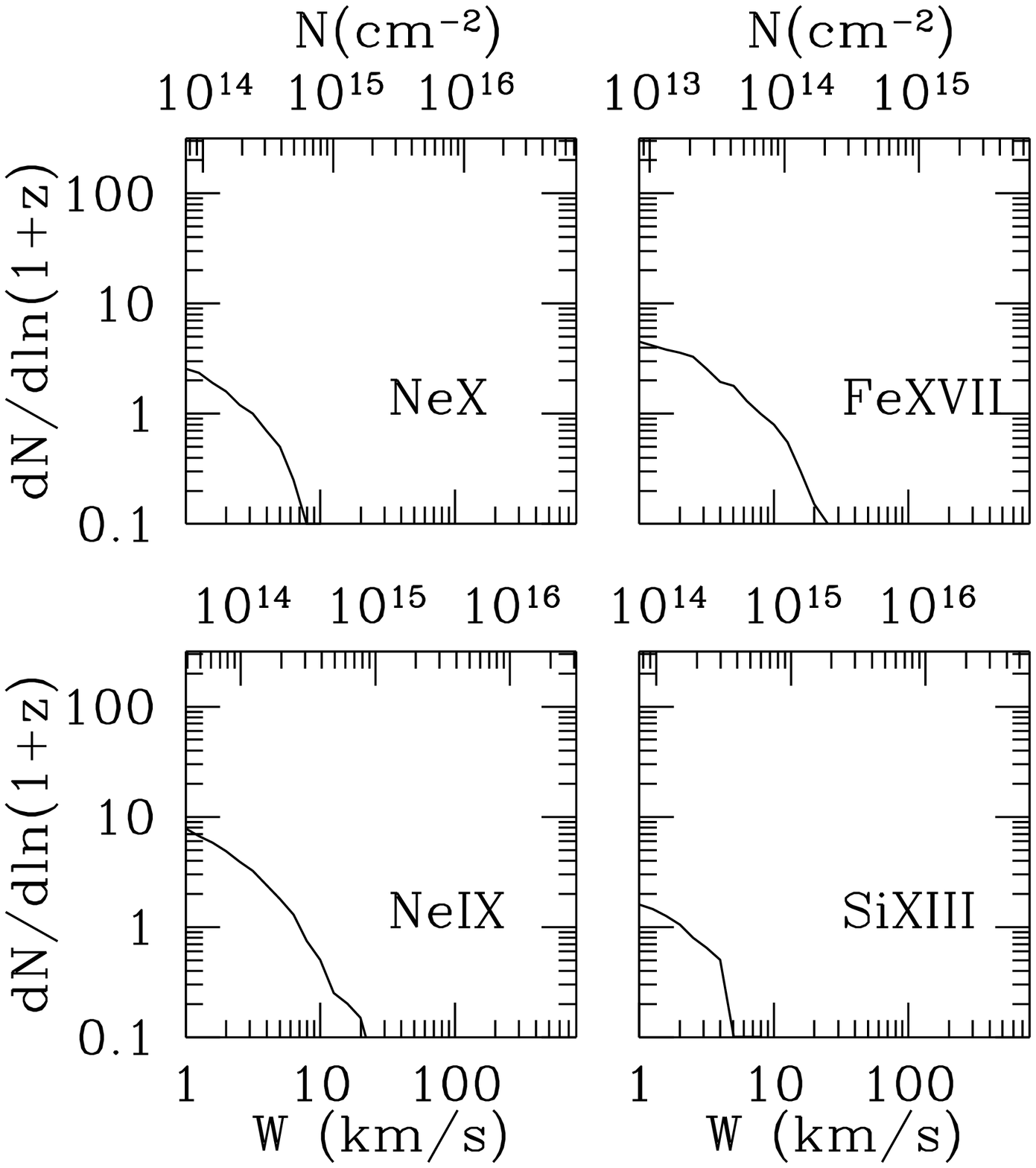}
}
\caption{
Like Fig.~\ref{optdistCN}, but for
\ion{Ne}{9}, \ion{Ne}{10},
\ion{Si}{13} and \ion{Fe}{17}.
}
\label{optdistNeSiFe}
\end{figure}

Our assumption of a uniform metallicity of $0.1Z_\odot$ with solar
abundance ratios (instead of the $\alpha$-enhanced ratios characteristic
of Type II SN enrichment) is probably conservative, and perhaps
unduly pessimistic.  The predicted equivalent width thresholds in
Table~\ref{ions} are approximately proportional to the assumed metallicity,
except that they grow more slowly for $W \ga 80\kms$ as saturation
becomes significant (see Figure~\ref{colO78}).
As shown in Figure~\ref{optdistO}, the predicted
number of systems at high equivalent width is sensitive to the
scatter in IGM metallicity as well as its mean value.  
For example, if we include 0.4 dex of scatter about 0.1 solar metallicity,
then the \ovii\ and \oviii\ equivalent width thresholds corresponding to
one line per unit redshift are about a factor of two higher than
the values quoted in Table~\ref{ions}.
Because the intrinsic equivalent width distribution is steep,
the first X-ray forest
systems to be detected are likely to lie on the high tail of the IGM
metallicity distribution.

We have devoted much of our attention to the physical state of the gas
producing \ovii\ and \oviii\ absorption, as illustrated in
Figures~\ref{zoombox}--\ref{O1-678}, especially Figure~\ref{Ophysprpty}.
Strong \ovii\ and \oviii\ absorbers, with $W > 50\kms$, arise mainly
in gas with $\delta \ga 100$ and temperature $T \sim 10^{5.5}\K - 10^{6.5}\K$,
where collisional ionization dominates over photoionization.
Weaker absorbers arise in lower density gas ($\delta \sim 30$ for
$W \sim 20\kms$) over a wider range of temperatures, with
photoionization often playing a significant role.
For $W>10\kms$, all \oviii\ absorbers and most \ovii\ absorbers
arise in the temperature regime $10^5\K < T < 10^7\K$ identified
by \cite{CO99a} as the ``warm-hot IGM,'' which contains a substantial
fraction of the low redshift baryons in cosmological simulations
\citep{CO99a,dave01}.  However, one should keep in mind that the
absorbers potentially detectable with {\it Chandra} or {\it XMM-Newton}
trace only the high density regions of the WHIM, with overdensities
characteristic of virialized groups rather than intergalactic filaments.

\begin{figure}
\begin{center}
\epsfig{file=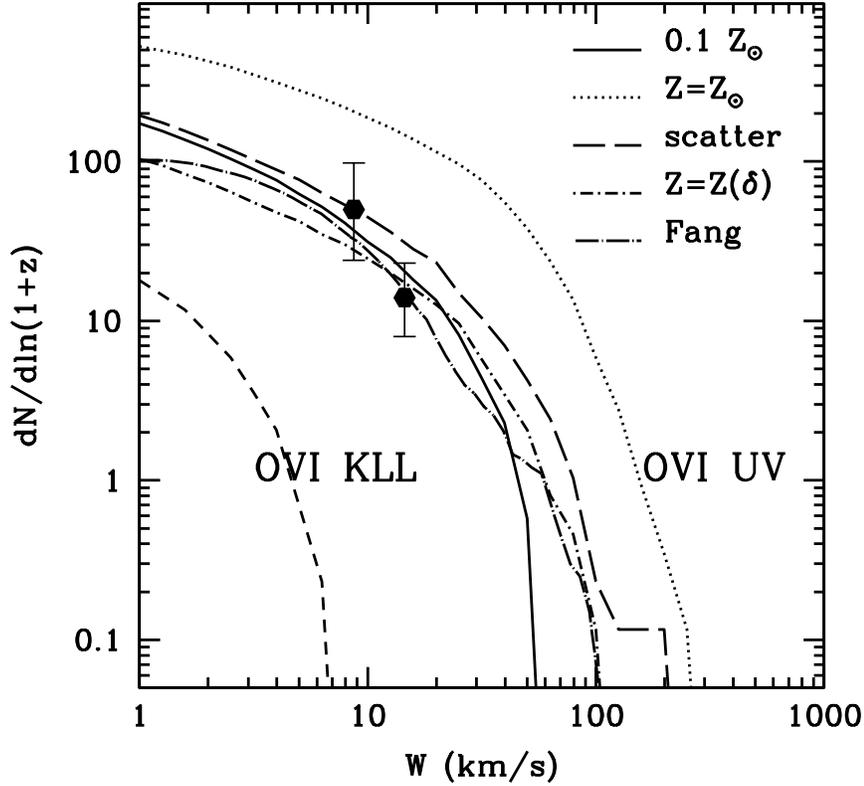,width=5in}
\end{center}
\caption{
Cumulative equivalent width distribution of \ovi\ absorbers,
with a range of metallicity assumptions,
all calculated for the Mi98+UV background.
Solid and dotted curves show predictions for the 1032\AA\ UV line,
assuming uniform metallicity $Z=0.1Z_\odot$ and $Z=Z_\odot$,
respectively.  Long-dashed and dot-short dashed curves show
calculations with, respectively, metallicity scatter and a combination of
metallicity scatter with density-dependent mean metallicity,
as in Figure~\ref{optdistO}.
The dot-long dashed curve shows the results of
\cite{fang01}, specifically their ``Model B,'' which incorporates
photoionization; their metallicity model is closest to that
of the $Z=Z(\delta)$ (dot-short dashed) curve above.
Points with error bars show corresponding observational data
from \cite{TSJ00} and \cite{S02},
as taken from figure 2 of \cite{Cen01}.
The short-dashed curve at lower left shows the equivalent width distribution
of KLL resonance absorption in the X-ray, for $Z=0.1Z_\odot$.
}
\label{O6}
\end{figure}

In agreement with \cite{Cen01} and \cite{fang01}, we find that the
$\Lambda$CDM model can plausibly account for the high incidence of
\ovi\ (1032\AA, 1038\AA) absorbers at low redshift
\citep{TS00,TSJ00,S02}, as shown in Figure~\ref{O6}.
The \ovi\ absorbers exhibit a well-defined trend between
equivalent width and overdensity, though there are some
extreme outliers (Figure~\ref{Ophysprpty}).  At every
equivalent width there is a substantial range of gas
temperatures (about a factor of ten), with photoionization
dominating in the cooler systems.  At $W \sim 20\kms$ ($\sim 70\mang$),
much of the \ovi\ absorption arises in gas with overdensity $\delta \sim 10$.

In searches with {\it Chandra} or {\it XMM-Newton}, \ovi\ absorbers
will play a crucial role as signposts for \ovii\ or \oviii\ absorption.
If the IGM metallicity is $0.1Z_\odot$, or even $0.3Z_\odot$, then
the predicted number of X-ray forest systems strong enough to be
detected at the $\sim 5\sigma$ level needed in a ``blind'' survey
is very small.  However, if one is searching at the redshift of
a known \ovi\ system, then a $2-3\sigma$ detection is already
significant, and the steepness of the equivalent width distribution
translates a small change in detection threshold to a large change in
predicted line density.  \ion{H}{1} Ly$\alpha$ absorption can also
serve as a signpost, but the higher density of Ly$\alpha$ lines and
the large scatter between \ion{H}{1} and \ovii\ or \oviii\
(Figure~\ref{O1-678}) makes it less useful for this purpose than \ovi.

Recent observations provide several tentative detections of intergalactic
\ovii\ and \oviii\ absorption.  The most convincing detections are those
of \cite{nicastro02}, who find absorption features of \ovii, \oviii, 
and \neix\ in their {\it Chandra} spectrum of PKS 2155-304, at wavelengths
consistent with zero redshift.  The implied \ovii\ and \oviii\ 
column densities are $\sim 5\times 10^{15}\cm^{-2}$.
{\it XMM-Newton} and {\it Chandra} observations also show $z=0$
absorption along lines of sight towards Mrk 421
and 3C 273 \citep{rasmussen03,fang03}.
These observations suggest that
highly ionized gas is fairly ubiquitous along lines of sight through
the Local Group, as one might expect when looking out from a local
maximum in the IGM density (see, e.g., \citealt{kravtsov02}),
though at least some of the absorption could be Galactic in origin
\citep{fang03}.
Our simulations predict that oxygen lines of this strength should
be rare along random lines of sight beyond the Local Group
(Fig.~\ref{optdistO}), and here the observational situation is
more ambiguous.  \cite{fang02b} find a statistically significant
absorption feature in the {\it Chandra} spectrum of PKS 2155-304
that could correspond to \oviii\ absorption coincident with a
known galaxy group at $cz=16,600\kms$, with an implied column 
density $\sim 10^{16}\cm^{-2}$.  However, the {\it XMM-Newton}
spectrum appears to rule out the existence of an absorption
feature at this level (Rasmussen, private communication; see
\citealt{rasmussen03}).
\cite{mathur03} use a {\it Chandra} spectrum of H1821+643 to search
for absorption at the redshifts of previously known \ovi\ 
systems \citep{TSJ00,Oeg00}, obtaining apparent detections
of one system in \ovii\ and another in both \ovii\ and \oviii.
The statistical significance of the features is only $\sim 2\sigma$
in each case, but because the absorption 
redshifts are known {\it a priori} from the \ovi\ measurements,
the probability that they are simply noise fluctuations is small ($\sim 5\%$).
The rest-frame equivalent widths are $\sim 150\kms$ and $\sim 110\kms$ for
the \ovii\ lines and $\sim 140\kms$ for the \oviii\ line, with
corresponding column densities $\sim 3-4 \times 10^{15}\cm^{-2}$ (\ovii)
and $\sim 7\times 10^{15}\cm^{-2}$ (\oviii), assuming no saturation.
The existence of two such strong systems in a path length 
$\Delta z \sim 0.3$ is surprising relative to our predictions
in Figure~\ref{optdistO}, unless the metallicity in these systems
is substantially higher than 0.1 solar, though it is worth noting
that this line of sight also has an unusually high incidence
of \ovi\ absorption \citep{Oeg00,TSJ00}.

Further 
measurements of or limits on \ovii\ and \oviii\ absorption associated
with \ovi\ will take us much further towards understanding this
newly discovered population of low redshift \ovi\ absorbers,
providing constraints on their densities, temperatures, and ionization
mechanisms and on their contribution to the cosmic baryon budget.
However, \ovi\ absorbers trace only the low temperature end of the
X-ray forest; many strong \ovii\ and \oviii\ systems have very weak
\ovi\ absorption, and, conversely, many strong \ovi\ systems will
be undetectable in \ovii\ or \oviii\ (Figure~\ref{Ocorr}).
A full accounting of the strong \ovii\ and \oviii\ absorbers will
probably require a mission with the capabilities of {\it Constellation-X},
which may also begin to detect X-ray absorption from
other elements in the same systems.  
Even at the $\sim 30\kms$ threshold of {\it Constellation-X},
the detectable X-ray forest absorbers at $Z=0.1Z_\odot$
correspond mainly to gas in the
outer regions of groups and poor clusters, with some contribution from
more diffuse gas in filaments.  The powerful {\it XEUS} satellite
would allow a comprehensive investigation of X-ray forest absorption
by a number of ionization species, providing great insight into
the distribution of shock-heated baryons and
the enrichment of the intergalactic medium.

\acknowledgements

We are grateful to Smita Mathur for helpful advice at
many stages in this work, especially in matters related to observational
capabilities.  We also thank Jordi Miralda-Escud\'e for several
valuable discussions and Joop Schaye for comments on the manuscript.
We thank Todd Tripp for providing the observational
data points used in Figure~\ref{O6} and Taotao Fang for providing
the numerical results from \cite{fang01} and FBC
(discussed in \S\ref{sec:ewdist} and shown in Figure~\ref{O6}).
This work was supported by NASA LTSA Grant NAG5-3525,
NSF Grant AST-9802568, and Chandra Observatory Grant G01-2118X from
Smithsonian Astrophysical Observatory.
X.C. was supported 
at OSU by the Department of Energy under grant DE-FG02-91ER40690, and
at ITP/UCSB by the NSF under grant PHY99-07949.
D.W. acknowledges the hospitality of the Institute for Advanced Study
and the financial support of the Ambrose Monell Foundation during
the completion of this work. R.D. was supported by NASA through
Hubble Fellowship grant number HST-HF-01128.01-A awarded by
the Space Telescope Science Institute, which is operated by AURA, Inc.,
under NASA contract NAS5-26555.
The simulation was performed at the San Diego Supercomputer Center
and NCSA.

\clearpage
%\onecolumn[

\begin{table}[t]
\begin{center}
\caption{\label{ions} \bf Ions and their strongest transition lines}
\begin{tabular}{llllll}
Ion & $E$ (keV) & $\lambda$ (\AA) & $f$ & $10^5 Z/H$ & $W_1 (\kms)$\\
\ion{C}{5} & 0.3079 & 40.27 & 0.65 & 35.5 & 18 \\
\ion{C}{6} & 0.3675 & 33.73 & 0.42 & 35.5 & 13 \\
\ion{N}{6} & 0.4307 & 28.79 & 0.68 & 9.33 & 5 \\
\ion{N}{7} & 0.5003 & 24.78 & 0.42 & 9.33 & 2.5 \\
\ion{O}{7} & 0.5740 & 21.60 & 0.70 & 74.1 & 40 \\
\ion{O}{8} & 0.6536 & 18.97 & 0.42 & 74.1 & 25 \\
\ion{Ne}{9} & 0.9220 & 13.45 & 0.72 & 11.7& 7 \\
\ion{Ne}{10} & 1.022 & 12.13 & 0.42 & 11.7 & 3 \\
\ion{Si}{13} & 1.865 & 6.648 & 0.76 & 3.55 & 2 \\
\ion{Fe}{17} & 0.8257 & 15.02 & 3.0 & 3.24 & 8 
\end{tabular}
\end{center}
Transitions producing the strongest X-ray forest absorption
in our simulation.  Columns 2-6 list the energy, wavelength, oscillator
strength, element abundance by number relative to hydrogen for
$Z=Z_\odot$, and the equivalent width threshold above which we find one
line per unit redshift assuming $Z=0.1Z_\odot$.  Atomic data and
abundances are from \cite{transitionlist}.
\end{table}

\begin{table}[h]
\caption{\label{instrument}
\bf Instrument properties: spectral resolution and effective area }
%\begin{tabular}{|l|r|r|r|r|r|r|}
\begin{tabular}{lrrrrrr}
%\tableline
 & & 0.3keV(41\AA) & 0.5keV(25\AA) &1keV(12\AA) & 2keV(6\AA) &  6keV(2\AA) \\
 & & & & & & \\
%\tableline
Chandra$^{1}$ & R & 1000 & 500 & 250 & 120 & 40\\
 & A & 8 & 25 &60 & 100 & $>100$\\
%\tableline
XMM-Newton$^2$ & R &800 & 500 &250 & 120 & --\\
 & A & 30  &70&  70 & 20 & -- \\
%\tableline
Constellation-X$^3$ &R & 800 & 500 & 300 & 1000 & 3000\\
 &A & 1000 & 3000 & 15000 & 9000 & 6000\\
%\tableline
XEUS$^4$ & R &700 & 800 & 1000 & 1000 & 1000\\
 &A & 30000 & 40000 & 40000 & 30000 & 30000\\
%\tableline
\end{tabular}

All effective areas are given in $\cm^2$, -- indicates no
response. \\
$^{1}$ using LETG ACIS, grating 1st order.\\
$^{2}$ using RGS,  grating 1st order.\\
$^3$ design goal.\\
$^4$ using STJ, design goal, initial mirror configuration.\\

\end{table}

\begin{table}[b]
\caption{\label{equivwidth}
\bf Equivalent Width Thresholds for $5\sigma$ Detection
}
\begin{tabular}{lrrrrrr}
%\tableline
 & & 0.3keV(41\AA) & 0.5keV(25\AA) &1keV(12\AA) & 2keV(6\AA) &  6keV(2\AA) \\
 & & & & & & \\
%\tableline
Chandra & eV & 0.24 & 0.45 & 1.3 & 4.7 & $<72$\\
 & km/s & 240 & 270 & 390 & 700 & $<2600$\\
 & m\AA & 33 & 22 &16 & 15 & $< 18$\\

%\tableline
XMM-Newton & eV &0.14 & 0.27 &1.2 & 10 & -\\
 & km/s & 140  &160&  360 & 1600 &-  \\
& m\AA & 19  &13 &  15 & 32 & - \\
%\tableline
Constellation-X &eV & 0.024 & 0.041 & 0.076 & 0.17 & 0.76\\
 &km/s &24 &25 & 23 &26 & 38\\
 &m\AA &3.3 &2.0 & 0.94 &0.53 & 0.26\\
%\tableline
XEUS& eV &0.0047 & 0.0089 & 0.025 & 0.094& 0.59\\
 & km/s & 4.7 & 5.3 & 7.6 & 14 & 29\\
 & m\AA & 0.64 & 0.44 & 0.31 & 0.29 & 0.20\\
%\tableline
\end{tabular}
Equivalent widths corresponding to $S/N=5$, computed
from eq.~(\ref{eqn:wmin}) using the resolution and effective
area values of Table~\ref{instrument}, a 500 ksec integration time,
and a quasar spectrum
$F=2\times 10^{-3} (E/\keV)^{-2.35}
  \,{\rm photons}\,\cm^{-2}\,{\rm s}^{-1}\,\keV^{-1}$
similar to that of the $z=0.297$ quasar H1821+643.
\end{table}

\clearpage

\begin{figure}
\begin{center}

%\begin{minipage}{0.497\linewidth}
%\epsfig{file=Tmap_part.ps,width=\linewidth}
%\end{minipage}\hfill
%\begin{minipage}{0.497\linewidth}
%\epsfig{file=denmap_part.ps,width=\linewidth}
%\end{minipage}\hfill

%\begin{minipage}{0.497\linewidth}
%\epsfig{file=O7frac_part.ps,width=\linewidth}
%\end{minipage}\hfill
%\begin{minipage}{0.497\linewidth}
%\epsfig{file=O8frac_part.ps,width=\linewidth}
%\end{minipage}\hfill

\end{center}

\caption{
The gas particle distribution in a $25\hmpc \times 25\hmpc \times 25\hmpc$
subvolume of our simulation, with particles color-coded by
temperature (top left), density (top right), \ovii\ ion fraction (bottom left)
and \oviii\ ion fraction (bottom right).
The temperature scale runs from $T\sim 10^{3.5}\K$ (blue) through
$T\sim 10^5\K$ (red) to $T\sim 10^7\K$ (yellow).
The density scale runs from $\delta \sim 100$ (blue) through
$\delta \sim 5\times 10^3$ (red) to $\delta \sim 5\times 10^5$ (yellow).
The ion fraction scales are linear and run from $\sim 0.2$ (blue)
through $\sim 0.5$ (red) to $\sim 0.85$ (yellow).
% Commands were:
% xgas logtemp 3 8
% xgas logrho 0 5
% readarray 8_7_X1fracarray.out ; xarray 0.01 1 wrbb clipboth
% readarray 8_8_X1fracarray.out ; xarray 0.01 1 wrbb clipboth
}
\label{zoombox}
\end{figure}

\begin{figure}
\begin{center}
%\begin{minipage}{0.497\linewidth}
%\epsfig{file=ovii14-17c.ps,width=\linewidth}
%\end{minipage}\hfill
%\begin{minipage}{0.497\linewidth}
%\epsfig{file=ovii14.5-17c.ps,width=\linewidth}
%\end{minipage}\hfill

%\begin{minipage}{0.497\linewidth}
%\epsfig{file=ovii15-17c.ps,width=\linewidth}
%\end{minipage}\hfill
%\begin{minipage}{0.497\linewidth}
%\epsfig{file=ovii15.5-17c.ps,width=\linewidth}
%\end{minipage}\hfill

\end{center}
\caption{
The projected \ion{O}{7} column density through the $50\hmpc$ simulation
volume, which has a redshift depth $\Delta z = 0.0167$.
The maps have thresholds of $\log N_{\rm OVII}$ of
14 (top left), 14.5 (top right), 15 (bottom left), and
15.5 (bottom right), and are computed assuming a
metallicity $Z=0.1Z_\odot$.
The last three panels correspond roughly to the detection thresholds for
{\it XEUS}, {\it Constellation-X}, and {\it Chandra}/{\it XMM-Newton}.
For an assumed metallicity of $Z=0.3Z_\odot$, column densities
would increase by 0.5 dex, so a similar identification would hold
for the first three panels instead of the last three.
}
\label{boxmap}
\end{figure}

\begin{figure}
\centerline{
\epsfxsize=5.5truein
\epsfbox{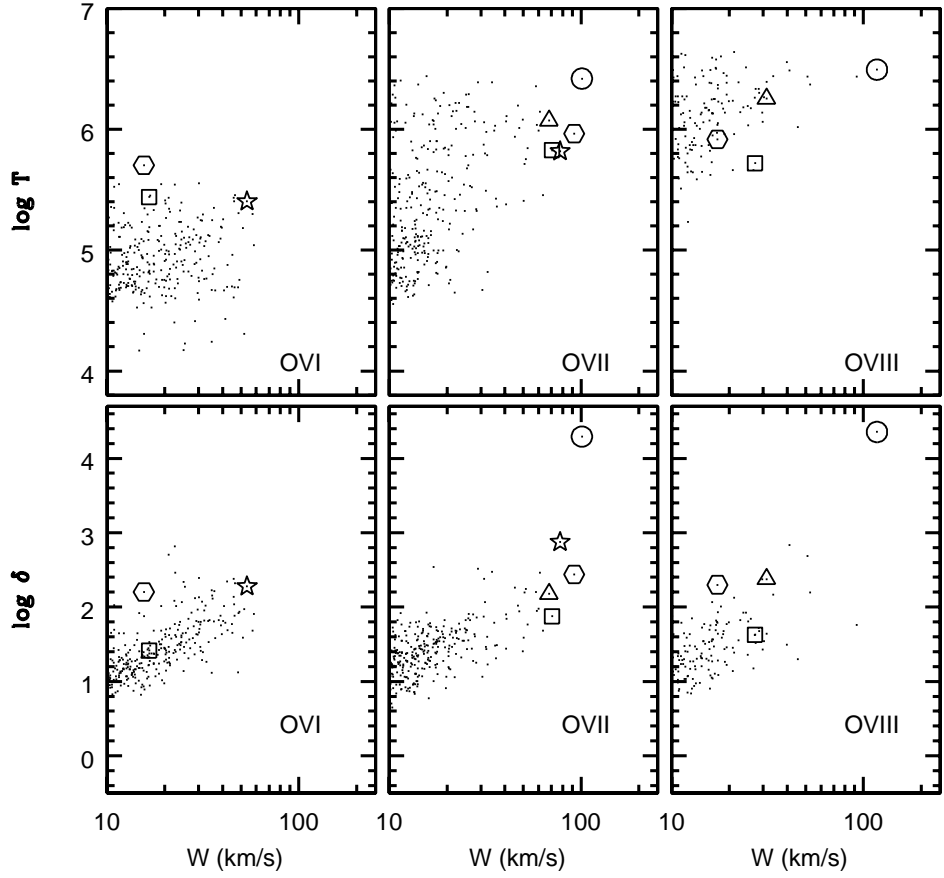}
}
\caption{
Temperature (top) and overdensity (bottom) of the gas associated
with \ovi, \ovii, and \oviii\ absorbers, plotted against absorber
equivalent width, for $Z=0.1Z_\odot$.
In this figure and the two that follow, the five systems with the
strongest \ovii\ absorption are marked with special symbols in each panel.
}
\label{Ophysprpty}
\end{figure}

\begin{figure}
\centerline{
\epsfxsize=2.5truein
\epsfbox{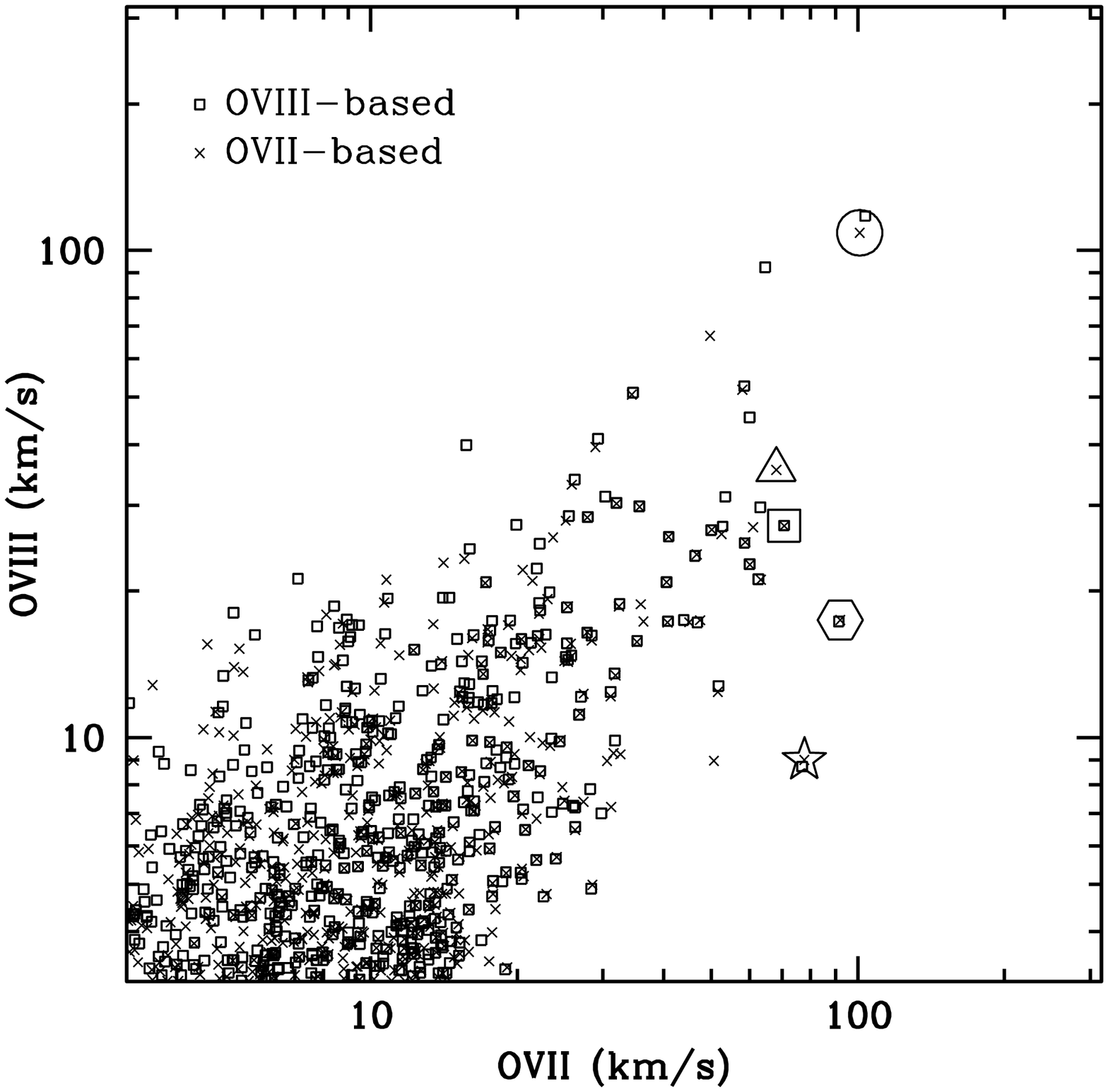}
}
\centerline{
\epsfxsize=2.5truein
\epsfbox{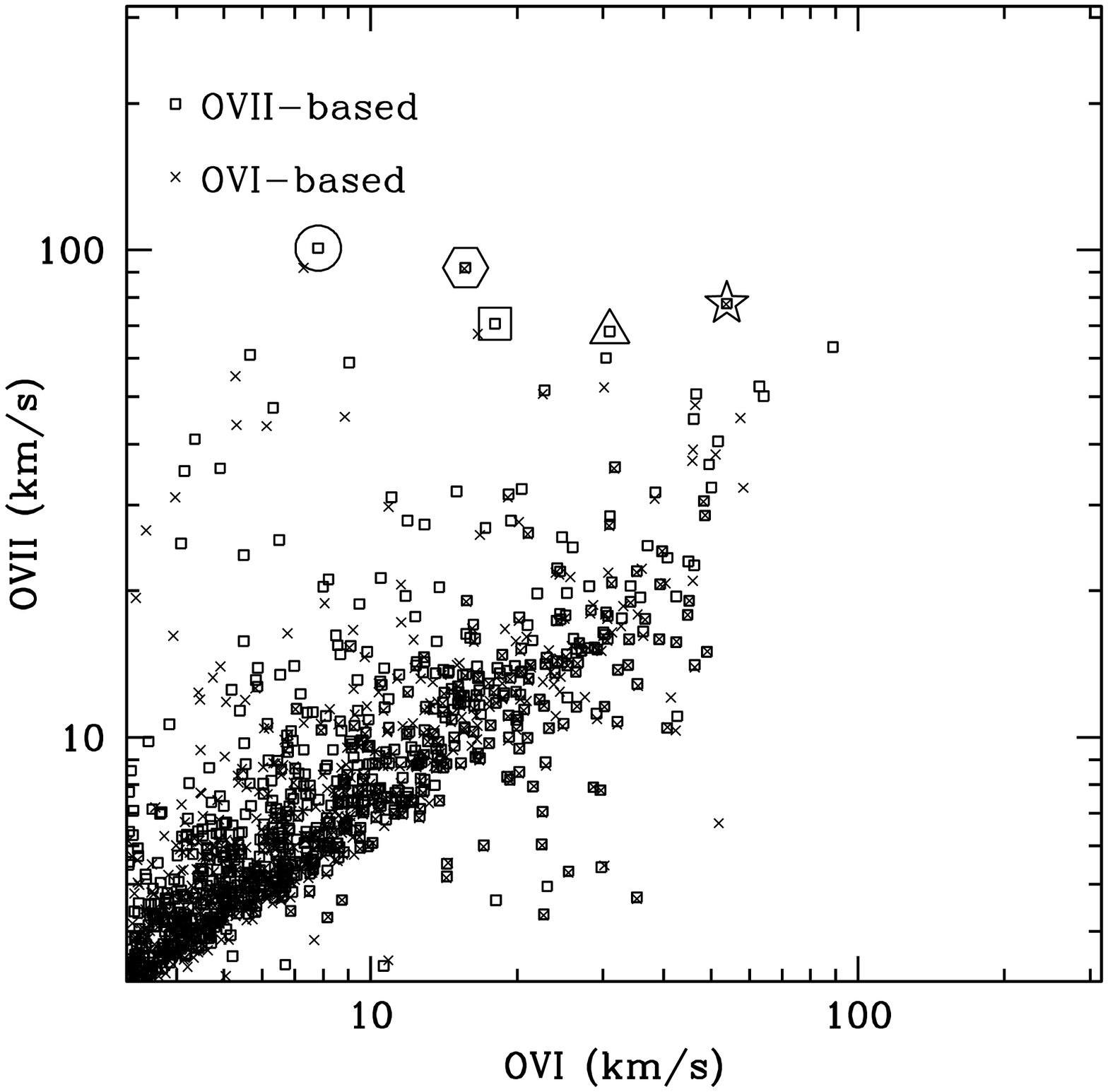}
}
\centerline{
\epsfxsize=2.5truein
\epsfbox{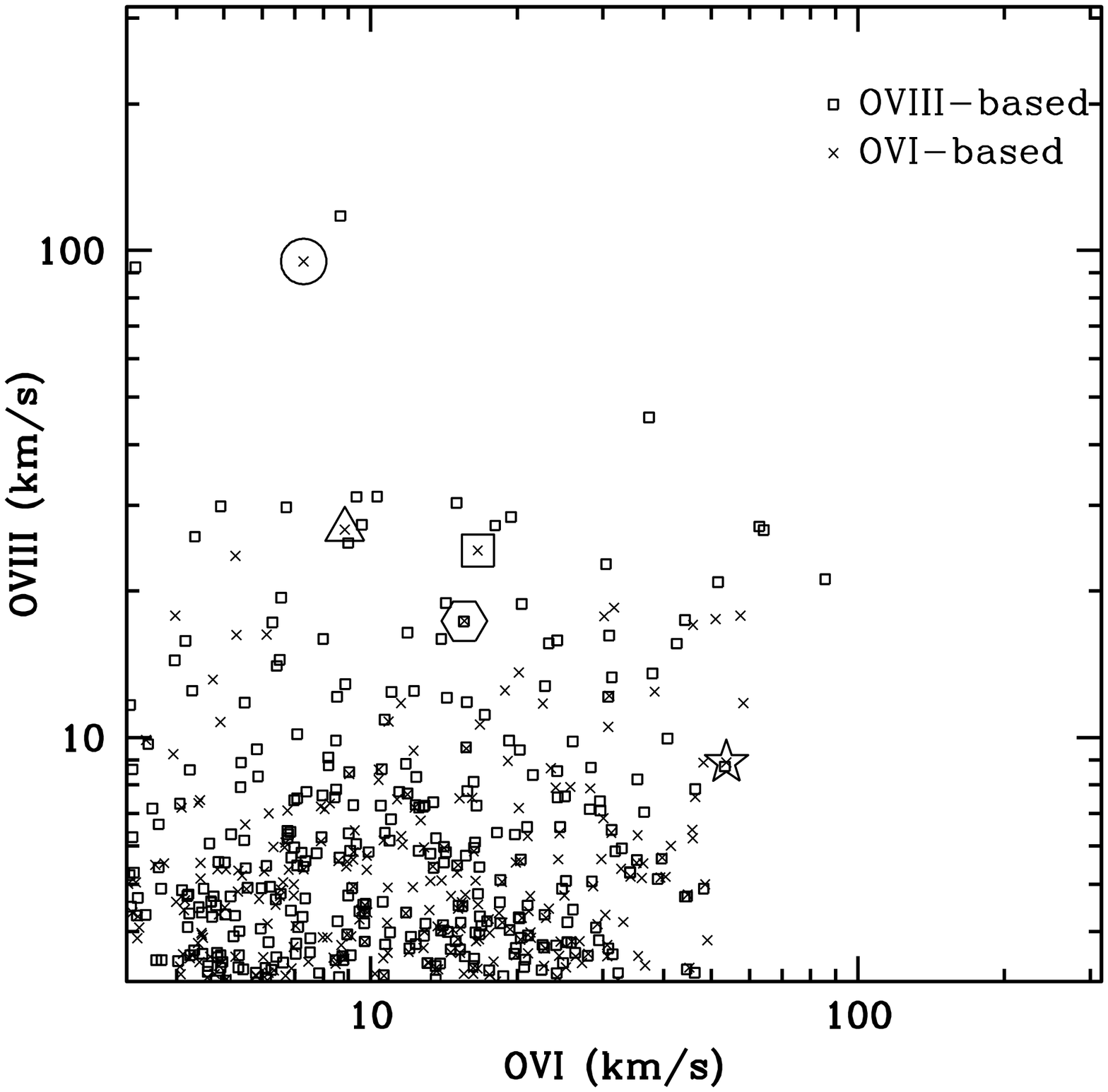}
}
\caption{
The equivalent width correlation for oxygen ions:
\ion{O}{7}--\ion{O}{8} (top),  \ion{O}{6}--\ion{O}{7} (middle),
\ion{O}{6}--\ion{O}{8} (bottom). Different symbols denote
absorber boundaries defined by the ion on the abscissa (crosses)
or ordinate (squares).
Large symbols mark the five systems with the strongest \ovii\ absorption.
}
\label{Ocorr}
\end{figure}

\begin{figure}
\centerline{
\epsfxsize=3.5truein
\epsfbox{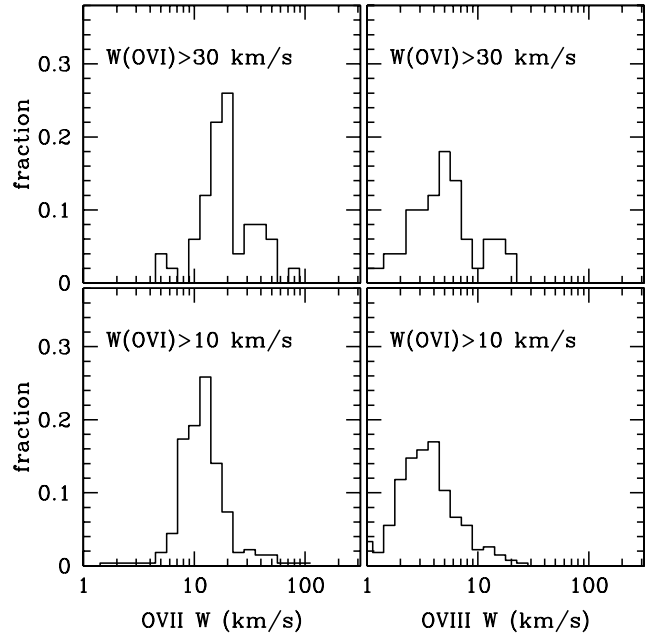}
}
\caption{
Histogram of \ion{O}{7} (left) and 
\ion{O}{8} (right) equivalent widths for \ion{O}{6} absorbers with 
W(\ion{O}{6}) greater than 30 km/s (top) and 10 km/s (bottom).
}
\label{myhist}
\end{figure}

\begin{figure}
\centerline{
\epsfxsize=3.5truein
\epsfbox{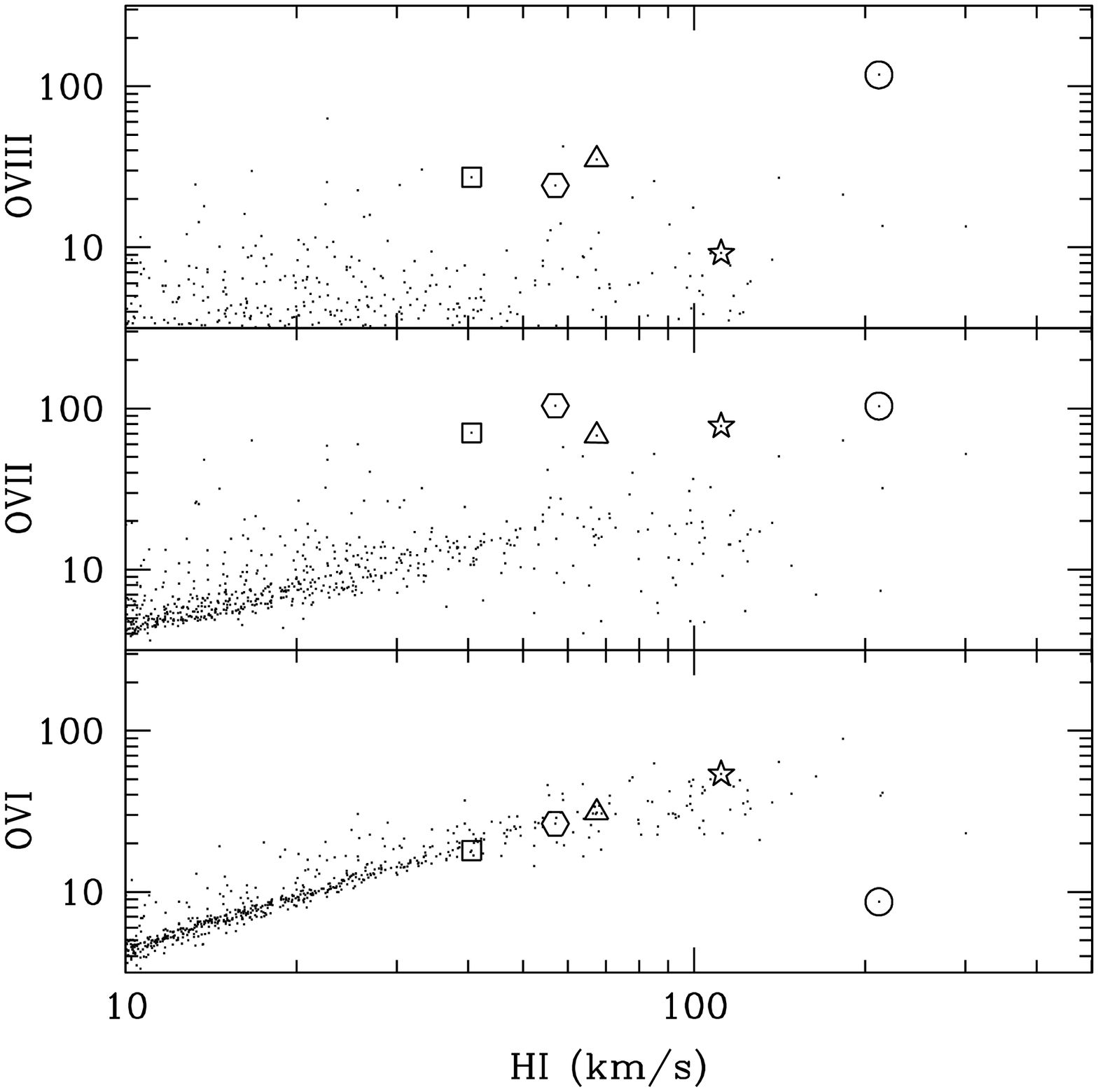}
}
\caption{
The correlation between
\ion{H}{1} and \ion{O}{6}, \ion{O}{7}, and \ion{O}{8} equivalent
width.
}
\label{O1-678}
\end{figure}

\end{document}